\documentclass[12pt]{article}
\usepackage{amsmath}
\usepackage{graphicx}
\usepackage[round]{natbib}
\usepackage{url} 
\usepackage[latin1]{inputenc} 
\newtheorem{theorem}{Theorem}
\newtheorem{corollary}{Corollary}
\newtheorem{lemma}{Lemma}

\newtheorem{remark}{Remark}
\newtheorem{proposition}{Proposition}

\newtheorem{definition}{Definition}
\newtheorem{assumption}{Assumption}

\newcommand\argmin{\operatorname{arg\,min}}

\newcommand\diag{\operatorname{diag}}
\usepackage{amssymb}
\usepackage{bm}
\usepackage{algorithmic}
\usepackage{booktabs}
\usepackage{amsmath}
\usepackage{algorithm}
\usepackage{algorithmic}
\renewcommand{\algorithmicrequire}{\textbf{Input:}}    \renewcommand{\algorithmicensure}{\textbf{Output:}}
\usepackage{color}
\usepackage{algorithm}
\usepackage{verbatim}
\usepackage{algorithmic}
\usepackage{makecell}
\usepackage{enumerate}
\usepackage{graphicx}
\usepackage{threeparttable}
\usepackage{multirow}
\usepackage{mathrsfs}

\usepackage{subfigure}
\usepackage{setspace}
\usepackage{xr-hyper}
\usepackage{hyperref}

\makeatletter
\newcommand*{\addFileDependency}[1]{
	\typeout{(#1)}
	\@addtofilelist{#1}
	\IfFileExists{#1}{}{\typeout{No file #1.}}
}
\makeatother

\newcommand*{\myexternaldocument}[1]{
	\externaldocument{#1}
	\addFileDependency{#1.tex}
	\addFileDependency{#1.aux}
}

\myexternaldocument{ppDSC_JCGS_SM}

 \hypersetup{
     colorlinks=true,
     linkcolor=magenta,
     filecolor=blue,
     citecolor = blue,
     urlcolor=cyan,
     }
\newcommand{\T}{{\intercal}}
\newcommand{\tA}{{\tilde A}}

\newcommand{\MP}{{\mathbb P}}
\newcommand{\ME}{{\mathbb E}}

\newcommand*{\QEDA}{\hfill\ensuremath{\blacksquare}}

\newcommand{\blind}{0}

\definecolor{darkgreen}{rgb}{0,0.6,0.2}

\begin{document}

\def\spacingset#1{\renewcommand{\baselinestretch}%
{#1}\small\normalsize} \spacingset{1}


\if0\blind
{
  \title{\bf Privacy-Preserving Community Detection for Locally Distributed Multiple Networks\thanks{Xiangyu Chang (xiangyuchang@xjtu.edu.cn) and Shujie Ma (shujie.ma@ucr.edu) are corresponding authors.}}
  \author{ Xiao Guo$^1$, Xiang Li$^2$, Xiangyu Chang$^3$, Shujie Ma$^4$   \hspace{.2cm}  \\
    $1$ School of Mathematics, Northwest University, China\\
    $2$ Department of Biostatistics, Epidemiology and Informatics \\ University of Pennsylvania, U.S.A. \\
    $3$  School of Management, Xi'an Jiaotong University, China\\
    $4$ Department of Statistics, University of California-Riverside, U.S.A. }
  \maketitle
} \fi

\if1\blind
{
  \bigskip
  \bigskip
  \bigskip
  \begin{center}
    {\LARGE\bf Privacy-Preserving Community Detection for Locally Distributed Multiple Networks}
\end{center}
  \medskip
} \fi

\bigskip
\begin{abstract}

This paper proposes a new efficient and scalable consensus community detection approach for multi-layer stochastic block models in a federated learning framework, where network layers are locally stored across several machines and have privacy-preserving needs. 
Specifically, we develop a spectral clustering-based algorithm named \textbf{{p}}rivacy-{\textbf{{p}}}reserving \textbf{{D}}istributed \textbf{{{S}}}pectral \textbf{{{C}}}lustering (\texttt{ppDSC}).
To reduce the bias incurred by the \emph{randomized response} (RR) mechanism for achieving \emph{differential privacy}, we develop a two-step bias adjustment procedure. For federated learning, we perform the eigen-decomposition of each privacy-preserved network locally and then aggregate the local eigenvectors using orthogonal Procrustes transformation on the central server.
We establish a novel bound on the misclassification rate of \texttt{ppDSC}. The new bound reveals the asymmetric roles of the two edge-flipping probabilities of the RR in the misclassification rate. Through the bound, we can also find the optimal choices for the flipping probabilities given a fixed privacy budget. Moreover, we show that \texttt{ppDSC} enjoys the same statistical error rate as its centralized counterpart when the number of machines satisfies a polynomial order with the sample size on each local machine and the \emph{effective heterogeneity} is well controlled.

\end{abstract}


\vfill

\newpage
\spacingset{1.5}

\setlength{\abovedisplayskip}{3pt}
\setlength{\belowdisplayskip}{3pt}

\section{Introduction}
\label{sec::intro}

Advances in modern technology have facilitated the collection of multiple network data from various sources (e.g., Facebook, LinkedIn, and Twitter) that present disparate social ties \citep{kairouz2021advances}.
This type of data can be well represented by multi-layer networks \citep{mucha2010community}, where the nodes are the entities of interest, and the network layers reflect their relationships across different local sources. In practice, a multi-layer network is often stored in a distributed manner because of ownership and communication costs, among others \citep{kairouz2021advances}. Moreover, the network edges from local
sources often contain sensitive and private information~\citep{dwork2014algorithmic}. Therefore, the analysis of the multi-layer networks should be naturally based on a distributed and federated architecture with additional privacy-preserving guarantees.

In the literature, the multi-layer stochastic block model (multi-layer SBM) \citep{han2015consistent,paul2016consistent,barbillon2017stochastic} has been popularly used for the community detection task in multi-layer networks, in which each layer is assumed to share the same unobserved community membership. Among the various methods for community detection under multi-layer SBMs, spectral clustering-based methods stand out because of their computational tractability.
There are two 
{\color{black} problems in} developing the spectral clustering-based method for multi-layer SBMs in a privacy-preserving and distributed learning paradigm. The first {\color{black}problem} is how to obtain theoretically sound statistics for each layer of networks in a privacy-preserving way. The randomized response (RR) mechanism \citep{warner1965randomized,karwa2017sharing} that simply flips all edges in the network with a certain probability has been developed to achieve the rigorous \emph{differential privacy} (DP) notion \citep{dwork2006calibrating}.
However, it undoubtedly brings bias into the multi-layer SBMs. Thus, a bias-correction procedure is crucially needed for community detection in the composite and RR-perturbed multi-layer SBMs.

The second {\color{black}problem} is how to develop a communication-efficient distributed learning algorithm for detecting communities under multi-layer SBMs. Recent years have seen various spectral clustering-based methods developed for community detection in multi-layer networks, say, the methods based on the aggregated adjacency matrix~\citep[e.g.,][]{paul2020spectral,lei2020bias} and the tensor-based methods~\citep[e.g.,][]{jing2021community}.
However, these methods are developed in the centralized paradigm, where all network layers are stored on a single central machine. In the federated learning paradigm, directly sending the RR-perturbed networks stored on the local machines to the central server can bring enormous communication costs or storage costs to the server, especially for large networks. Therefore, performing the eigen-decomposition step of spectral clustering on each local machine is {a natural way} to reduce the communication cost.
A follow-up problem is how to combine the eigenvectors to cope with subspaces' orthogonal ambiguity.

In this paper, we propose a \textbf{{p}}rivacy-{\textbf{{p}}}reserving \textbf{{D}}istributed \textbf{{{S}}}pectral \textbf{{{C}}}lustering (\texttt{ppDSC}) algorithm with only one-round of communication for detecting the communities of locally stored multi-layer network. In particular, we assume that the layers of an $L$-layer network are distributed across $m\ (m\leq L)$ machines. In our algorithm, each machine first performs a carefully designed two-step bias-correction procedure on the \emph{squared} RR-perturbed local networks, where we adopt the {{squared}} matrix to avoid possible cancellation of communities among different layers \citep{lei2020bias}. Then, the eigen-decomposition is performed locally. Finally, the central server collects the local eigenspaces and conducts the spectral clustering using the average of \emph{orthogonal Procrustes transformed} (OPT) \citep{schonemann1966generalized,cape2020orthogonal,charisopoulos2021communication} eigenspaces.


The main contributions of this paper are summarized below. First, we propose a one-shot-based distributed spectral clustering algorithm for detecting the communities of RR-perturbed multi-layer SBMs. In particular, we develop a debiased estimator to accommodate the noise incurred by the RR mechanism and the square transformation. The debiased estimator can consistently estimate the original population matrix, which leads to smaller eigenvector and misclassification error bounds than the non-debiased estimator (with respect to its population matrix or the original population matrix). To the best of our knowledge, the theoretical examination of the bias-correction is rarely addressed in existing literature. We propose to aggregate the local eigenspaces instead of the debiased matrices for communication efficiency. Due to orthogonal ambiguity, we propose aligning the local eigenspaces with OPT (\citet{cape2020orthogonal,charisopoulos2021communication}).
It is worth mentioning that the algorithm is general because it includes the special cases where the number of local machines is one or the privacy-preserving constraint disappears.

Second, we provide novel bounds of the distributed algorithm \texttt{ppDSC} regarding the eigenvector estimation and clustering, which explicitly show how the 
{\color{black}misclassification rate} depends on the privacy parameters. In particular, when the privacy budget in DP increases logarithmically with the number of nodes, the error bounds coincide with those without the privacy-preserving constraint.
Two interesting findings are observed from the privacy-aware bounds. First, without a pre-specified privacy budget, perturbing the network entries from 0 to 1 is more harmful than perturbing the network from 1 to 0 to the upper error bound. {\color{black} Second, with a given privacy budget, from our theoretical bounds, one can find the optimal edge-flipping probabilities in the RR mechanism.
To the best of our knowledge, these new theoretical results for RR-perturbed DP algorithms have not been investigated in the SBM literature.}
In addition, the conditions for the bounds reveal that denser networks can allow a more strict privacy-preserving level, i.e., a smaller privacy budget.

Last but not least, the bounds show how local computation and heterogeneity affect the performance of \texttt{ppDSC}. Note that in the regression set-up, the one-shot-based distributed algorithm can attain the centralized bound {\color{black}when the number of machines satisfies a polynomial order of the sample size on each local machine; }
see \citet{zhang13communication,lee2017communication}, among many others. Similarly, we show that the error bound for our \texttt{ppDSC} decomposes into a term that corresponds to the centralized error and two remainder terms corresponding to the \emph{squared local error} and the \emph{effective heterogeneity}. The two remaining terms are negligible when the number of machines satisfies a condition similar to the one in the regression setting, and the effective heterogeneity is not severe. Then, the estimator converges at the same rate as a centralized estimator. Our work is the first to derive an error bound for a privacy-preserving and distributed learning algorithm in multi-layer SBMs.

\subsection{Related work}
\label{subsec:relatedwork}
Recent years have seen increasing work on multi-layer SBMs including \citet{han2015consistent,paul2016consistent,lei2020consistent,bhaskara2019distributed,paul2020spectral,arroyo2021inference,lei2020bias,jing2021community}, among many others. A common assumption in these works is that all network layers are centrally stored {without considering privacy concerns.} As a result, 
{the community detection is performed directly on the original data on the central server. Different from the aforementioned works, we consider the federated learning setting in which the network layers are stored locally on each machine with privacy preservation need.}
 To improve communication efficiency, we propose to work with the locally computed eigenvectors instead of the original data, which results in both methodological and theoretical differences between our methods and the aforementioned works.

There have been various DP algorithms developed for analyzing network data, including graph statistics releasing \citep{nissim2007smooth,karwa2011private,imola2021locally}, synthetic graphs releasing \citep{qin2017generating}, generative graph models estimation \citep{mir2012differentially,borgs2015private,karwa2016inference,chang2021edge}, community detection \citep{nguyen2016detecting,ji2019differentially} among others. These algorithms are not developed based on SBMs. Under the single SBM, two recent works \citet{hehir2021consistency} and \citet{seif2022differentially} studied respectively the misclassification rate and information-theoretical limits of community detection under the RR mechanism. The differences between our work and these two works are as follows. First, we study the multi-layer SBM while they studied the single SBM. Second, we derive several new privacy-aware bounds that reflect the asymmetric roles of the two edge-flipping probabilities in RR and indicate the best choice of the edge-flipping probabilities given a fixed privacy budget. These observations are new to the literature. In addition, we theoretically show the effect of bias-correction, which was previously unclear.

Driven by large decentralized datasets, there has been increasing work on distributed algorithms for PCA, including the divide-and-conquer paradigm \citep{fan2019distributed,bhaskara2019distributed,charisopoulos2021communication} and iterative paradigm \citep{garber2017communication,gang2019fast,chen2021distributed,guo2021privacy}. However, we work with {privacy-preserved} network data and incorporate the perturbation and bias-correction procedure in our algorithm; as a result, it poses great challenges to derive the statistical theories, including the estimation error bounds.

\subsection{Notation}
{We use $\lambda_{\min}(A)$ to denote the smallest non-zero eigenvalue of a given matrix $A$. Moreover, we use the projection distance $\mathrm{dist}(U,U'):= \|UU^\T-U'U'^\T\|_2$
to measure the subspace distance between any pair of orthogonal basis $U,U'\in \mathbb R^{n\times K}$. $\|A\|_2$ denotes the spectral norm of the matrix $A$ or the Euclidean norm of vector $A$ and $\|A\|_{\tiny F}$ denotes the Frobenius norm of matrix $A$. In addition, we use the following standard notation for asymptotics. We write $f(n)\asymp g(n)$ or $f(n)=\Theta(g(n))$ if $c g(n)\leq f(n)\leq C g(n)$ for some constants $0<c<C<\infty$, $f(n)\lesssim g(n)$ or $f(n)=O(g(n))$ if $f(n)\leq Cg(n)$ for some constant $C<\infty$, and $f(n)\gtrsim g(n)$ or $f(n)=\Omega(g(n))$ if $f(n)\geq cg(n)$ for some constant $c>0$. Finally, the constants throughout this work may be different from place to place.

\subsection{Organizations}

The rest of the paper is organized as follows. Section \ref{sec::problem} provides the preliminaries of the model and DP. Section \ref{sec::alg} develops the algorithm for privacy-preserving community detection of locally stored networks. Section \ref{sec::theory} studies the theoretical properties of \texttt{ppDSC}. Section \ref{sec:sim} and \ref{sec::real} include the simulation and real data experiments. Section \ref{sec::concl} concludes the paper.

\section{Problem formulation}
\label{sec::problem}

In this section, we first present the multi-layer SBMs for generating the locally stored networks. Next, we provide the formal definition of the DP and RR mechanisms.

\subsection{The model for locally stored networks}

Suppose there are $L$ aligned networks with $n$ common nodes. The networks are binary and symmetric with the adjacency matrix denoted by $A_l\in \{0,1\}^{n\times n},\;l\in[L]:=\{1,2,\dots,L\}$. In the multi-layer  SBMs \citep{han2015consistent,paul2016consistent}, each $A_{l}$ is independently generated as follows. The $n$ nodes are assigned to $K$ non-overlapping communities with the community assignment of node $i$ denoted by $g_i\in[K]$. Given the community assignments, each entry $A_{l,ij}\ (i<j)$ of $A_l$ is generated independently according to
\begin{equation}
\label{MOD:SBM}
A_{l,ij}\sim {\mathrm{Bernoulli}} (B_{l,g_ig_j}),\quad i<j, \;l\in[L],
\end{equation}
where $B_l$'s are connectivity matrices, and the diagonal entries $A_{l,ii}$'s are all 0. The heterogeneity among the networks is inherent in $B_l$'s, and we will quantify it in the next section.

We assume $A_{l\in[L]}$'s are locally generated and stored at $m$ different machines. For notational simplicity, we assume each machine evenly owns $s:=L/m$ networks. We use $S_i:=\{(i-1)s+1,..., i\cdot s\}$ to denote the index set of networks stored in the $i$th machine.

Given the generation process of $A_l$'s, we consider the RR mechanism for the adjacency matrices for private protection in the context of DP \citep{dwork2014algorithmic}.

\subsection{Differential privacy and randomized response}
Though the local networks $A_{l\in S_i}$'s are not directly shared by all the local machines, potential privacy breaches still exist \citep{dwork2017exposed}. DP provides a rigorous framework for privacy-preserving data analysis \citep{dwork2006calibrating}. Generally, a DP algorithm pursues that its output appears similar in probability if one individual changes the input data. By definition, DP protects any individual's privacy from an adversary who has access to the algorithm output and even sometimes the rest of the data. In addition, it can prevent other kinds of attacks, say, the re-identification attack, the tracing attack, and the membership attack, among others; see \citet{dwork2017exposed} for a survey. In the context of networks, we consider the following notion of edge-DP \citep{rastogi2009relationship,karwa2017sharing}.

\begin{definition}[Edge-DP]
\label{edp}
Let $A\in \{0,1\}^{n\times n}$ be the network whose edges are sensitive. For any range $\mathcal R$, let $\mathcal Q(A)\in \mathcal R$ be a randomized mechanism that generates some synthetic information from $A$, indicated by a known parameter $\omega$. Then, $\mathcal Q(A)$ is said to be $\epsilon$-edge-DP if
\begin{equation}
\label{dpdef}
\underset{\tA\in \mathcal R}\max \,\underset{A',A,\, \Delta(A,A')=1}\max\; \log \frac{\mathbb P_{\omega}(\mathcal Q(A)=\tA\,| \,A)}{\mathbb P_{\omega}(\mathcal Q(A')=\tA \,|\,A')}\leq \epsilon,
\end{equation}
where $\Delta(A,A')=1$ means that $A$ and $A'$ are neighboring by differing only one edge.
\end{definition}

\begin{remark}
We make the following remarks on the DP.
\item[(1)] Edge-DP protects the privacy of network edges by ensuring that the algorithm's output does not reveal the inclusion or removal of any certain edge in the network. In other words, edge-DP can protect against an adversary who knows the algorithm's output and the existence or absence of edges between all pairs but one pair of nodes.
\item[(2)] The $\epsilon$ is often termed as the privacy budget. A smaller privacy budget indicates stronger protection of privacy. Note that DP is the worst-case definition, and the randomness in \eqref{dpdef} only comes from the DP mechanism $\mathcal Q$.
\item[(3)] Like the general DP, edge-DP enjoys the post-processing property~\citep{dwork2014algorithmic}, that is, an $\epsilon$-edge-DP algorithm is still $\epsilon$-edge-DP after any post process that does not use the knowledge about the original network.
\end{remark}

Edge-DP could be achieved by flipping the network edges with certain probabilities, which is called the randomized response (RR) mechanism \citep{warner1965randomized,karwa2017sharing,ayed2020information}. The algorithm outputs a perturbed adjacency matrix $\tilde{A}$ of $A$. Specifically, for each $i<j$, when $A_{ij}=1$, $\tilde{A}_{ij}$ remains 1 with probability $q$ and flips to 0 otherwise. Similarly, when $A_{ij}=0$, $\tA_{ij}$ remains 0 with probability $q'$ and changes to 1 otherwise. The diagonal entries $\tA_{ii}=0$. Note that $q$ and $q'$ can be published publicly. The relationship between $q,q'$ and the privacy budget $\epsilon$ is shown in the following proposition.

\begin{proposition}[\citet{karwa2017sharing}]
\label{dp-prop}
The randomized response mechanism satisfies $\epsilon$-edge-DP with
\begin{equation}
\label{pribudg}
\epsilon\geq \log\max \left\{ \frac{q'}{1-q},\frac{1-q}{q'},\frac{1-q'}{q},\frac{q}{1-q'}\right\}.
\end{equation}
\end{proposition}
According to \eqref{pribudg}, we can easily derive that
for a given $\epsilon$,
\begin{equation*}
\max\{1-\mathrm{e}^\epsilon q,\, \frac{1-q}{\mathrm{e}^\epsilon}\}\leq q'\leq \min\{\mathrm{e}^\epsilon (1-q),\, 1-\frac{q}{\mathrm{e}^\epsilon}\}.
\end{equation*}
\textcolor{black}{
We plot the feasible region of $q$ and $q'$ for a given $\epsilon$ in Figure \ref{region}, where the formulation of the four boundaries is shown in red. For fixed $\epsilon$, larger $q$ and $q'$ will lead to less accuracy loss. Hence, the upper boundaries, i.e., $q'=1-\frac{q}{\mathrm{e}^\epsilon}$ when $q\leq \frac{\mathrm{e}^\epsilon}{1+\mathrm{e}^\epsilon}$ and $q'=\mathrm{e}^\epsilon (1-q)$ when $q\geq \frac{\mathrm{e}^\epsilon}{1+\mathrm{e}^\epsilon}$, are more desirable. A typical choice of $q$ and $q'$ is that
\begin{equation}
\label{eq}
q=q'=\frac{\mathrm{e}^\epsilon}{1+\mathrm{e}^\epsilon},
\end{equation}
which has been proven to be optimal in maximizing the mutual information between raw and RR-perturbed data \citep{ayed2020information}. More importantly, it turns out that this choice is also optimal in the context of this work in the sense that it leads to the smallest misclassification upper bound; see discussions after Theorem \ref{centralbound}.
}

\begin{figure}[h]{}
\centering
{\includegraphics[height=8cm,width=8cm,angle=0]{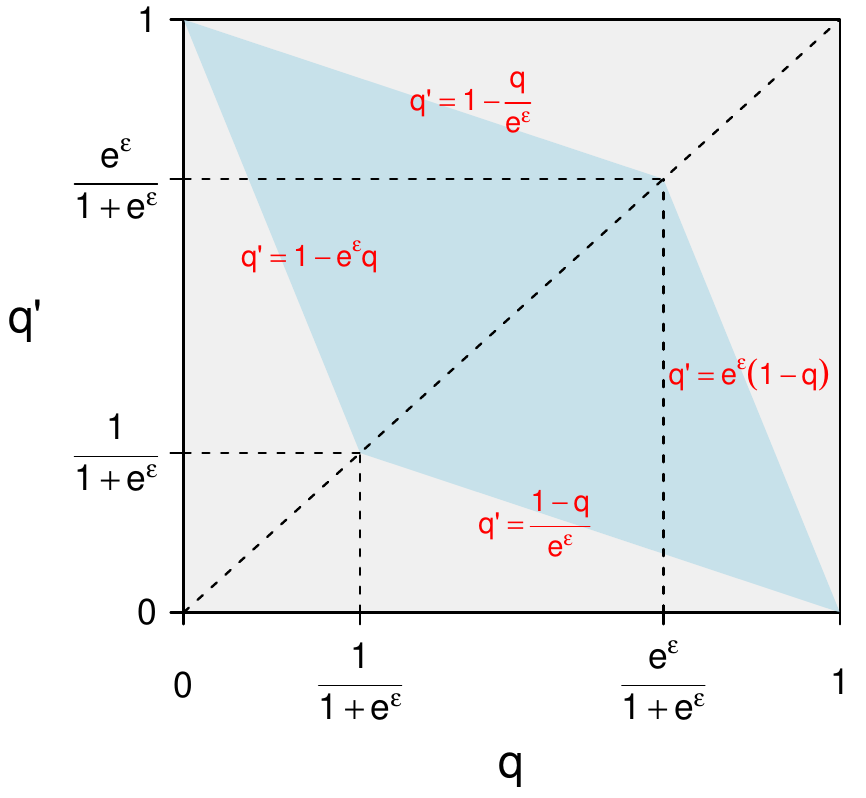}}
\caption{The feasible region (in blue) of $q$ and $q'$ under a given privacy budget $\epsilon$.}\label{region}
\end{figure}

Although RR can achieve edge-DP, one can imagine that the resulting network largely deviates from the original network. In the next section, we will see how we can de-bias the perturbed networks to improve the algorithm's utility under the multi-layer SBMs.

\section{Privacy-preserving community detection for locally distributed networks}
\label{sec::alg}
This section proposes the privacy-preserving community detection algorithm for locally distributed networks.
In particular, we develop a bias-adjustment procedure to remove the bias inherent in the RR-perturbed networks.

Recall that there are $m$ local machines, and the $i$th machine owns $s$ adjacency matrices $A_{l\in S_i}$'s. $A_{l\in [L]}$'s are assumed to be generated from a multi-layer SBM. Denote the population of $A_l$ by $P_l:=\Theta B_l\Theta^\T$, where $\Theta\in \mathbb R^{n\times K}$ is the membership matrix which is formed by forcing $\Theta_{ig_i}=1$ and $\Theta_{ij}=0$ for $j\neq g_i$. We aim to recover the common membership matrix $\Theta$ using the network information.

Since $P_l$'s are not necessarily positive semidefinite, the top-$K$ singular vectors of $P_l$'s are a good choice to reveal the communities with both assortative and disassortative structures \citep{rohe2011spectral}, which is equivalent to the top-$K$ eigenvectors of $P_l^2$. In the following, we will develop a bias-adjustment procedure to obtain a good estimator for $P_l^2$ when we only have access to the RR-perturbed adjacency matrices.

\subsection{Bias-adjustment}
Our debiasing procedure involves two steps: in step 1, we build a debiased estimator for the connectivity matrix upon the RR-perturbed adjacency matrix stored in each local machine; in step 2, we further remove the bias of the diagonals caused by taking the square of the debiased estimated matrix obtained in step 1.

{Step 1.} We now illustrate the necessity of bias adjustment when dealing with RR-perturbed data.

For $1\leq l\leq L$, we denote the RR-perturbed $A_l$ by $\tA_l$. Specifically, for $i<j$, if $A_{l,ij}=1$,
\begin{equation}
\label{rr1}
\tA_{l,ij}=\begin{cases}
1, \quad \mathrm{with\, probability}\, q,\\
0, \quad \mathrm{otherwise},\end{cases}
\end{equation}
and $\tA_{l,ji}=\tA_{l,ij}$.
On the other hand, if $A_{l,ij}=0$,
\begin{equation}
\label{rr2}
\tA_{l,ij}=\begin{cases}
1, \quad \mathrm{with\, probability}\, 1-q',\\
0, \quad \mathrm{otherwise},\end{cases}
\end{equation}
and $\tA_{l,ji}=\tA_{l,ij}$.
By (\ref{rr1}) and (\ref{rr2}), taking expectations for two times, we have
\begin{equation*}
\mathbb E(\tA_{l,ij})=P_{l,ij}(q+q'-1)+1-q',
\end{equation*}
which is biased from the true expectation $P_{l,ij}$ of $A_{l,ij}$. Therefore, for $i\neq j$, the local machine only needs to de-bias $\tA_{l,ij}$ to be
$$\overline{A}_{l,ij}=\frac{\tA_{l,ij}-(1-q')}{q+q'-1}.$$ We thus obtain the following debiased estimator $\overline{A}_l$ for $P_l$,
\begin{equation}\label{firstdb}
\overline{A}_l:=[\tA_l-(1-q')\mathbf 1\mathbf 1^\T+\mathrm{diag}(1-q')]/(q+q'-1),
\end{equation}
where $\mathbf 1$ denotes the vector with all entries being 1.

{Step 2.} Though $\overline{A}_l$ is the unbiased estimator for $P_l$, the diagonal elements of $\overline{A}_l^2$ can cause bias for $P_l^2$. It turns out that the error would be smaller if we further subtract $\frac{\sum_{j}\tA_{l,ij}(q')^2}{(q+q'-1)^2}$ from the $i$th diagonal elements of $\overline{A}_l^2$ provided that $\max_{i,j,l}P_{l,ij}=o(1)$. In particular, this debiasing step is crucial when $L$ is increasing. To save space, the details are relegated to Section \ref{app::biasadj} of Supplementary Materials.


Combining the two steps of debiasing, we eventually obtain the following estimator $\overline{M}_l$ for $P_l^2/n$,
\begin{equation}
\label{secdb}
\overline{M}_l:= \frac{\overline{A}_l^2}{n}-\frac{(q')^2}{n(q+q'-1)^2}G_l,
\end{equation}
where $G_l$ is a diagonal matrix with $G_{l,ii}=\sum_{j}\tA_{l,ij}$, and $\overline{A}_l$ and $\tA_l$ are defined in \eqref{rr1}-\eqref{firstdb}. Note that the second step of debiasing is in a similar spirit of \citet{lei2020bias} except that {their debiasing is performed on the original network matrix, while ours is on the debiased RR-perturbed network matrix.  }


\subsection{Privacy-preserving distributed spectral clustering}
Since DP enjoys the post-processing property, in principle, each local machine can send the averaged estimator $\sum_{l\in S_i}\overline{M}_l/s$ to the central server where the spectral clustering is further performed on the aggregated matrix
\begin{equation}
\label{barM}
\overline{M}:=\frac{1}{m}\sum_{i=1}^m\sum_{l\in S_i} \frac{\overline{M}_l}{s}.
\end{equation}
However, both the communication and computational cost can be very high for sending the whole dense matrices $\sum_{l\in S_i}\overline{M}_l/s$'s and performing further aggregation of these matrices on the central server, especially when the matrix dimensions are large. Instead, we consider a more efficient distributed algorithm, as described below.

We adopt the one-shot averaging framework in \citet{charisopoulos2021communication} for distributed computing performed on our debiased RR-perturbed network matrices. That is, on the $i$th machine, we first compute the eigen-decomposition on $\sum_{l\in S_i}\overline{M}_l/s$, and obtain the matrix $\hat{V}_i\in\mathbb R^{n\times K}$ that consists of the first $K$ eigenvectors, for all $i\in [m]$. On the central server, we average all the local solutions.
Note that direct averaging is inferior due to the orthogonal ambiguity of the problem, i.e., for any orthogonal matrix $Z$, $\hat{V}_iZ$ also forms a valid orthonormal basis of the subspaces formed by $\hat{V}_i$. Hence, $\hat{V}_i$'s are not guaranteed to be close to each other, even if their corresponding subspaces are close. In light of this, the local $\hat{V}_i$'s must be aligned with each other before averaging.

To avoid the orthogonal ambiguity of the problem, we use the \emph{Procrustes fixing} strategy \citep{schonemann1966generalized,cape2020orthogonal,charisopoulos2021communication,guo2021privacy}. For two matrices $V_1,V_2$ to be aligned, we can map $V_2$ to $V_1$ using an orthogonal transformation solved by the following optimization,
\begin{equation}
\label{opt}
Z:= \argmin _{Z\in \mathcal O_K} \| V_1-V_2Z\|_{F},
\end{equation}
{where $\mathcal O_k$ denotes the set of $K\times K$ orthogonal matrices}.
Note that (\ref{opt}) has a closed-form solution. Let the SVD of $V_2^\T V_1$ be $LDR^\T$, then $Z:=LR^\T$ is the solution to (\ref{opt}). In such a way, we can align local eigenspaces $\hat{V}_i$ \textcolor{black}{to an arbitrarily chosen reference eigenspace, for example, $\hat{V}_o$}, after which the aligned eigenspaces can then be averaged and orthogonalized. \textcolor{black}{As we will see in Section \ref{sec::theory}, the selection of the reference eigenspace does not affect the theoretical bounds.}

\begin{algorithm}[!htbp]
\small

\renewcommand{\algorithmicrequire}{\textbf{Input:}}

\renewcommand\algorithmicensure {\textbf{Output:} }

\caption{\textbf{P}rivacy-\textbf{p}reserving \textbf{d}istributed \textbf{s}pectral \textbf{c}lustering (\texttt{ppDSC})}

\label{Oneshotsvd}

\begin{algorithmic}[1]

\STATE \textbf{Input:} Network adjacency matrices $\{A_l\}_{l=1}^L$, number of machines $m$, number of clusters $K$, privacy parameters $q,q'$, index of reference solution $o$.  \\
\STATE \textbf{{{The} local machines do:}}
\FOR{$i = 1$ to $m$}
\FOR{$l$ in $S_i$}
\STATE \texttt{RR-perturbation:} Perturb $A_l$ to obtain $\tA_l$ using \eqref{rr1} and \eqref{rr2};\\
\STATE \label{debiasline}\texttt{Bias-correction:} Constructs the debiased estimator $\overline{M}_l$ according to (\ref{secdb});\\
\ENDFOR
\STATE \texttt{Eigen-decomposition:} Performs eigen-decomposition on $\sum_{l\in S_i}\overline{M}_l/s$ and denote the resulting eigen-spaces as $\hat{V}_i$.
\ENDFOR
\STATE \textbf{{{The} central server does:}}
\STATE \texttt{Procrustes fixing:} $\tilde{V}_i:=\hat{V}_iZ_i$ with $Z_i:= \argmin _{Z\in \mathcal O_K} \|\hat{V}_iZ- \hat{V}_o\|_{F}$ for each $i=1,...,m$;\\
\STATE \texttt{Aggregation:} Aggregates the local solutions according to $\bar{V}:= \frac{1}{m}\sum_{i=1}^m \tilde{V}_i$;\\
\STATE \texttt{Orthogonalization:} Performs the QR decomposition to obtain $\tilde{V},\tilde{R}={\rm qr} (\bar{V})$.
\STATE \texttt{Clustering:} Conduct $k$-means clustering on $\tilde{V}$ with cluster number being $K$. The resulting clusters are formulated by the membership matrix $\tilde{\Theta}$.\\
\STATE \textbf{Output:} The approximated eigen-space $\tilde{V}$ and the estimated membership matrix $\tilde{\Theta}$.

\end{algorithmic}
\end{algorithm}

The proposed \textbf{p}rivacy-\textbf{p}reserving \textbf{D}istributed \textbf{S}pectral \textbf{C}lustering (\texttt{ppDSC}) algorithm is summarized in Algorithm \ref{Oneshotsvd}. It should be noted that, in a real scenario, the computation on the local machine may be fulfilled by a single party or cooperated by different parties. For example, a party with data ownership may desire to obtain good utility while preserving the data's privacy. Thus, all the local steps are conducted within the party. It can also happen that
the RR perturbation is conducted by the data collection department of a company, while the data analysis department could do the remaining bias-correction and eigen-decomposition steps.

Algorithm \ref{Oneshotsvd} can achieve strong privacy protection in the DP's framework.
\begin{theorem}
\label{privacy-theory}
Define $A_{\mathrm{all}}:=(A_1,...,A_L)$. Denote the output of Algorithm \ref{Oneshotsvd} by $\mathcal M\in \mathcal R$. For $\epsilon$ satisfies (\ref{pribudg}), Algorithm \ref{Oneshotsvd} is $\epsilon$-edge-DP in the following sense
\begin{equation*}
\underset{\mathcal M\in \mathcal R}\max \,\underset{A_{\mathrm{all}},A_{\mathrm{all}}',\ \Delta(A_{\mathrm{all}},A_{\mathrm{all}}')=1}\max\; \log \frac{\mathbb P(\mathcal M(A_{\mathrm{all}})\,| \,A_{\mathrm{all}})}{\mathbb P(\mathcal M(A_{\mathrm{all}}') \,|\,A_{\mathrm{all}}')}\leq \epsilon,
\end{equation*}
where $\Delta(A_{\mathrm{all}},A_{\mathrm {all}}')=1$ means that all layers of the multi-layer network pairs $A_{\mathrm{all}}$ and $A_{\mathrm{all}}'$ are equal except one edge of one layer.
\end{theorem}

\section{Theoretical analysis}
\label{sec::theory}
In this section, we theoretically study the \texttt{ppDSC} algorithm in terms of the estimation accuracy of the eigenvectors and the misclassification rate.

To fix the ideas, we now recall and introduce some notations. We assume that there are $L$ networks with adjacency matrices being $A_l\in \{0,1\}^{n\times n}$'s evenly distributed on $m$ local machines; each machine owns $s$ networks. For $i\in [m]$, the $i$th machine has networks indexed by set $S_i$. The adjacency matrices $A_l$'s are generated from SBMs with layer-specific connectivity matrices $B_l$'s and a consensus membership matrix $\Theta\in \mathbb R^{n\times K}$. {We here assume that the number of communities $K$ is known. When $K$ is unknown, we will investigate the method and theory for the estimation of $K$, which is a model selection problem, as a future research topic, as this current work focuses on estimation and clustering analysis for a given model based on RR-perturbed network matrices.}
The number of nodes within community $k$ is denoted by $n_k$. The privacy parameters are denoted by $q,q'$ ($1/2<q,q'\leq 1$); see also \eqref{rr1} and \eqref{rr2}. Our goal is to use \texttt{ppDSC} (Algorithm \ref{Oneshotsvd}) to recover the first $K$ eigenspaces $V$ of the underlying target matrix $Q$ defined by
\begin{equation}
\label{q}
{Q}:=\frac{1}{L}\sum_{l=1}^L \frac{P_l^2}{n}:=\frac{1}{L}\sum_{l=1}^L {Q}_l,
\end{equation}
and the membership matrix $\Theta$.

The following assumptions and definition are generally required to derive the theoretical results for community detection.

\begin{assumption}[Balanced communities]
\label{balcom}
The community is balanced such that for any $k\in\{1,...,K\}$, $c\frac{n}{K}\leq n_k\leq C\frac{n}{K}$
for some constants $0\leq c\leq C$.
\end{assumption}

\begin{assumption}[Sparsity]
\label{sparse}
The connectivity matrix $B_l =\rho B_{l,0}$, where $\rho=o(1)$ and $B_{l,0}$ is a $K\times K$ symmetric matrix with constant entries in $[0,1]$.
\end{assumption}

\begin{assumption}[Full rank]
\label{rank}
$\sum_l B_{l,0}^2$ is of full rank $K$ with $\lambda_{{\min}}(\sum_l B_{l,0}^2)\geq cL$.
\end{assumption}

\begin{definition}[Heterogeneity]
\label{balb}
Suppose the connectivity matrix $B_l$'s are heterogeneous and define a heterogeneous measurement as
\begin{equation*}
\mathcal H(n, \rho, L):=\max_{i\in [m]} \|\frac{1}{s}\sum_{l\in S_i}B_l\Delta ^2B_l-\frac{1}{L}\sum_{l\in [L]}B_l\Delta^2 B_l\|_2,
\end{equation*}
where $\Delta:={\diag}(\sqrt{n_1},...,\sqrt{n_{K}})$.
\end{definition}

\begin{remark}
Assumption \ref{balcom} requires that the communities are balanced. Assumption \ref{sparse} specifies the connectivity matrices. In particular, we require the network sparsity $\rho=o(1)$. Assumption \ref{rank} requires the linear growth of the aggregated squared connectivity matrices in terms of its minimum eigenvalue, which is trivially satisfied when $B_{l,0}$'s are equal and is also used in \citet{lei2020bias}. Assumption \ref{rank} combining with Assumptions \ref{sparse} and \ref{balcom} implies that $\sum_{l=1}^L B_l \Delta^2B_l$ is of full rank, and $Q$ is of rank $K$; see the proof of Theorem \ref{sparsecor}. Indeed, the full rank requirement could be extended to the rank-deficient case; see \citet{zhang2022randomized} for example. Definition \ref{balb} measures how far the local average $\sum_{l\in S_i}B_l\Delta ^2B_l/s$'s deviates from their global average $\sum_{l\in [L]} B_l\Delta^2 B_l/L$.
In what follows, we call $\mathcal H$ the measure for the heterogeneity.
\end{remark}


We first provide the following error decomposition theorem to present the general error bound for the output eigen-space of \texttt{ppDSC}.

\begin{theorem}[Error Decomposition]
\label{oneshot-theory}
Suppose Assumption \ref{balcom} holds, then the
output $\tilde{V}$ of Algorithm \ref{Oneshotsvd} satisfies
\begin{equation*}
\mathrm{dist}(\tilde{V}, V)\lesssim {\max_{i\in[m]}\frac{\|\sum_{l\in S_i}(\overline{M}_l-Q_l)/s\|_2^2}{\lambda_{{\min}}^2 (Q)}}+{\frac{\mathcal H^2(n,\rho, L)}{\lambda_{{\min}}^2 (Q)}}+\frac{\|\sum_l\overline{M}_l/L-Q\|_2}{\lambda_{{\min}}(Q)}:=\mathcal I_1+\mathcal I_2+\mathcal I_3.
\end{equation*}
{provided that both $\mathcal I_1$ and $\mathcal I_2$ are smaller than a small constant $c$.}

\end{theorem}
\begin{remark}
{The condition that $\mathcal I_1$ and $\mathcal I_2$  are smaller than a small constant $c$  is a standard condition \citep{damle2020uniform,charisopoulos2021communication}; it ensures that the maximum error for the local average $\max_{i\in[m]}\|\sum_{l\in S_i}\overline{M}_l/s-Q\|_2$ is smaller than the eigen-gap $\lambda_{{\min}}(Q)$, which further indicates that the $K$th largest eigenvalues of $\sum_{l\in S_i}\overline{M}_l/s$ are separated from its remaining eigenvalues by Weyl's inequality. \textcolor{black}{In Theorem \ref{sparsecor}, we show that this condition is satisfied by our setting under mild conditions and further provide the explicit expressions for $\mathcal I_1$ and $\mathcal I_2$.}}

\textcolor{black}{The results of Theorem \ref{oneshot-theory} hold for any reference eigenvector $\hat{V}_o$ provided that we choose the true eigenspace $V$ to be aligned with $\hat{V}_o$ in the sense that
$\argmin _{Z\in \mathcal O_K} \| \hat{V}_oZ-V\|_{F}=\mathbb I_K$
by noting that we have a degree to choose the version of eigenspace $V$ \citep{charisopoulos2021communication}. Empirically, since it is easy to see that each local machine has the same population-wise eigenspaces, the experimental results would be similar when choosing a different sample-wise reference solution $\hat{V}_{o}$.}

\end{remark}

\begin{remark}
The decomposed three components $\mathcal I_1,\mathcal I_2,\mathcal I_3$ are explained as follows. $\mathcal I_1$ corresponds to the maximum squared error of the local estimators $\hat{V}_i$'s except that we use $Q$ in the denominator instead of $\sum_{l\in S_i}Q_l/s$'s; see the Davis-Kahan theorem on the perturbation of eigenvectors (\citet{davis1970rotation}; see Lemma \ref{lem:DK} in the Supplementary Materials). As a result, we only require the averaged population matrix $Q:=\sum Q_l/L$ be of rank $K$, and the local averaged population matrix $\sum_{l\in S_i}Q_l/s$ {can be} rank deficient, which is one of the benefits of combining the local eigenvectors. $\mathcal I_2$ corresponds to the effect of heterogeneity. To distinguish, we call $\mathcal H$ the measure for heterogeneity and call $\mathcal I_2$ the measure for \emph{effective heterogeneity}. Note that $\mathcal H$ and $\lambda_{{\min}} (Q)$ jointly decide $\mathcal I_2$. Therefore, if the local averages are noisy from the signal such that combining them across machines could hide the true signal, i.e., $\mathcal H$ is large, and $\lambda_{{\min}} (Q)$ is small, then the effective heterogeneity $\mathcal I_2$ would be large. On the other hand, if the local averages are far from the signal but combining them would enhance the signal, i.e., $\mathcal H$ is large, but $\lambda_{{\min}} (Q)$ can also be large such that their ratio is small, then the effective heterogeneity $\mathcal I_2$ would be small. {We refer to the two simulated examples in Section \ref{app::exp} of the Supplementary Materials for more illustrations on the aforementioned phenomena}.
$\mathcal I_1$ and $\mathcal I_2$ can be regarded as the price needed to pay for the distributed computation. Moreover, $\mathcal I_3$ corresponds to the error of the centralized estimator $\hat{V}:=\frac{1}{L}\sum_l\overline{M}_l$.
\end{remark}
Before developing our main result, namely, the specific version of Theorem \ref{oneshot-theory}, we first provide Theorem \ref{centralbound} to study how the privacy parameters $q,q'$ affect the upper bound for the centralized estimator $\|\sum_l\overline{M}_l/L-Q\|_2$.
We need the following definition of $\xi (q,q')$ and $C(q,q')$ through out the remainder of this section to simplify the notation,
\begin{equation}
\label{simplenote}
\xi (q,q')= \frac{1-q'}{(q+q'-1)^2}\quad {\rm and} \quad C(q,q')=\frac{q}{(q+q'-1)^2},
\end{equation}
where we note that when $q,q'>1/2+\tau$ with a small constant $\tau$, $C(q,q')$ can be regarded as a constant. We further use $\xi_{q,q'}$ and $C_{q,q'}$ to simplify the notation.

\begin{theorem}
\label{centralbound}
Suppose Assumption \ref{sparse} holds, and
\begin{equation}
\label{eandl}
\epsilon \leq \alpha_1\cdot {\log}\, n \quad {and}\quad  \log (L+n)\leq  n^{\alpha_2}
\end{equation}
for some constants $0<\alpha_1<1$ and $0<\alpha_2\leq 1-\alpha_1$,
the estimator $\overline{M}:=\frac{1}{L}\sum_l\overline{M}_l$ of $Q$ satisfies
\begin{align}
\label{bound1}
\small
\|\overline{M}-Q\|_2\lesssim\max &\bigg\{\frac{(C_{q,q'}\rho+\xi_{q,q'})^{1/2}n^{1/2}\rho \log^{1/2}(L+n)}{\sqrt{L}},\nonumber\\
&\quad\quad\quad\frac{(C_{q,q'}\rho+\xi_{q,q'})\log(L+n)}{\sqrt{L}},\, {(\rho+(q+q'-1)\xi_{q,q'})^2}\bigg\},
\end{align}
with probability larger than $1-O((L+n)^{-\nu})$ for some constant $\nu>0$.
\end{theorem}

\begin{remark}
Condition (\ref{eandl}) implicitly assumes that $q,q'$ relate to the privacy budget $\epsilon$ via (\ref{eq}). Nonetheless, the proof of Theorem \ref{centralbound} indicates that the result (\ref{bound1}) also holds under (\ref{pribudg}), i.e., the more general region of $q,q'$, with a modification of condition (\ref{eandl}). Theorem \ref{centralbound} provides the error bound for the centralized estimator $\sum_l\overline{M}_l/L$. When the privacy parameter $q'$ is large in the sense that $\xi_{q,q'}\lesssim C_{q,q'}\rho$, the upper bound in (\ref{bound1}) matches that in \citet{lei2020bias} up to log terms, where they did not impose the privacy-preserving constraint.
\end{remark}

Similar to Theorem \ref{centralbound}, we can easily derive the upper bound for $\max_i\|\sum_{l\in S_i}(\overline{M}_l-Q_l)/s\|_2$. Putting together the results, we provide in the following Theorem \ref{sparsecor} the specific error bound for the estimated eigen-spaces by Algorithm \ref{Oneshotsvd}.

It is well-known that large network density will help to recover the communities. Therefore, in what follows, we focus on the more challenging regime with $n{\rho^2}/(C_{q,q'}\rho+\xi_{q,q'})\lesssim {{\log (s+n)}}$, though our results can also be extended to $n{\rho^2}/(C_{q,q'}\rho+\xi_{q,q'})\gtrsim {{\log (s+n)}}$.

\begin{theorem}
\label{sparsecor}
Suppose Assumptions \ref{balcom}-\ref{rank} and condition (\ref{eandl}) hold.
If $n{\rho^2}/(C_{q,q'}\rho+\xi_{q,q'})\lesssim {{\log (s+n)}}$,
\begin{equation}
\label{rhosparse0}
n\sqrt{s}\cdot\frac{\rho^2}{C_{q,q'}\rho+\xi_{q,q'}}\gtrsim {{\log (s+n)}};
\end{equation}
\begin{equation}
\label{n}
(q+q'-1)\cdot\frac{\xi_{q,q'}}{\rho}\lesssim n^{1/2},
\end{equation}
then the
output $\tilde{V}$ of Algorithm \ref{Oneshotsvd} satisfies
\begin{equation*}
\mathrm{dist} (\tilde{V}, V)\lesssim \mathcal E_1+\mathcal E_2+\mathcal E_3;
\end{equation*}
\begin{equation*}
\mathcal E_1:= \max \bigg\{\frac{\log^2(s+n)(C_{q,q'}\rho+\xi_{q,q'})^2}{n^2{s}\rho^4},\;
\frac{(1+(q+q'-1)\xi_{q,q'}/\rho)^4}{n^2}
\bigg\},\; \mathcal E_2:=\frac{\mathcal H^2(n,\rho, L)}{n^2\rho^4};
\end{equation*}
\begin{equation*}
\mathcal  E_3:=\max \bigg\{\frac{\log(L+n)(C_{q,q'}\rho+\xi_{q,q'})}{n{\sqrt{L}}\rho^2}, \;\frac{(1+(q+q'-1)\xi_{q,q'}/\rho)^2}{n}
\bigg\}
\end{equation*}
with probability larger than $1-O((L+n)^{-\nu})-O(m(s+n)^{-\nu}))$ for some constant $\nu>0$.
\end{theorem}

In Theorem \ref{sparsecor}, $\mathcal E_1,\mathcal E_2$ and $\mathcal E_3$ correspond to the squared local error, effective heterogeneity, and centralized error, respectively. The first two errors are the price to pay for the unavailability of
full network layers. Condition (\ref{rhosparse0}) and (\ref{n}) are required to make the results in Theorem \ref{oneshot-theory} valid, which ensures $\mathcal I_1\lesssim 1$ in Theorem \ref{oneshot-theory}.    $\mathcal I_2\lesssim 1$ is automatically met as we have shown in the proof.  Several remarks are in order to explain the conditions and results in Theorem \ref{sparsecor}.

\begin{remark}[Effect of network sparsity]
Condition (\ref{rhosparse0}) on $\rho$ depends on the number of network layers $s$ in each local machine.
It reduces to $L^{1/2}n\rho\gtrsim \log (L+n)$ when $m=1$ and $\xi_{q,q'}\lesssim C_{q,q'}\rho$, which is identical to the condition in Theorem 1 of \citet{lei2020bias} up to log terms.
It is worth mentioning that the squared local error $\mathcal E_1$ and the follow-up condition on $\rho$ is inherent in
the distributed learning algorithm with only \emph{one} round of communication. See \citet{zhang13communication,lee2017communication} for example in the context of distributed empirical risk minimization and distributed sparse regression.  In the context of SBMs, as far as we are aware, this work is the first result to obtain a regression-like result for the one-shot-based distributed algorithm. It is also worth mentioning that \cite{arroyo2021inference} studied community detection on multi-layer networks, with each layer distributively stored in a separate machine. The requirement on $\rho$ turns out to be $\Omega(\log n/n)$, independent of network layers, where $n$ denotes the number of nodes. The condition on $\rho$ may be alleviated if we consider \emph{multi-round} communications as in the context of regression; see \citet{jordan2018communication,chen2020distributed}, among others. The power method-based approach provides a possible basis \citet{guo2021privacy}, which we left as our future work.

Condition (\ref{rhosparse0}) and (\ref{n}) both reveal that denser networks (i.e., larger $\rho$) can allow a more strict privacy-preserving level (i.e., small $\epsilon$). To see this, suppose $q,q'$ relate to $\epsilon$ via (\ref{eq}), then (\ref{rhosparse0}) reduces to
$\gamma(\epsilon):=\frac{(\mathrm{e}^\epsilon-1)^2}{\mathrm{e}^\epsilon+1}\gtrsim \frac{1}{\rho^2}\cdot \frac{\log(s+n)}{{ ns^{1/2}}}$,
where $\gamma(\epsilon)$ increases with $\epsilon$; (\ref{n}) reduces to $\mathrm{e}^\epsilon-1\gtrsim {1}/{(\rho n^{1/2})}$.
\end{remark}

\begin{remark}[Asymmetric roles of $(q,q')$]
From the definition of $C_{q,q'}$ and $\xi_{q,q'}$, it is undoubtedly that in general, the error bound of the eigenspace decreases when $q$ and $q'$ values increase.  Interestingly, we find that without a pre-specified privacy budget $\epsilon$, $q$ and $q'$ play an asymmetric role in the error bound when the
network is sparse with $\rho=o(1)$. The asymmetric role can be observed through the term $C_{q,q'}\rho+\xi_{q,q'}$. When $q,q'>1/2+\tau$ for an arbitrarily small constant $\tau>0$, $C_{q,q'}$ is a constant and $\xi_{q,q'}\asymp 1-q'$. Thus, $C_{q,q'}\rho+\xi_{q,q'}\asymp \rho+ 1-q'$, so we see that larger $1-q'$ would make the error bound larger, i.e., perturbing 0 to 1 is more harmful to the upper bound than perturbing 1 to 0. This interesting phenomenon can be explained as follows. We see that the sparsity level of the RR-perturbed network has the order of $\rho+ 1-q'$, so perturbing 1 to 0 does not change the order of the network sparsity so it does not affect the error bound with a large sample. However, perturbing 0 to 1 can make the network denser if $1-q'>\rho$, but the inflated ``1" values are the wrong and harmful ones, making the error bound worse.

\end{remark}

\begin{remark}[Optimal choice of $(q,q')$]
With a pre-specified specified privacy budget $\epsilon$, Theorem \ref{sparsecor} indicates that $q=q'=\frac{\mathrm{e}^\epsilon}{1+\mathrm{e}^\epsilon}$  is the best choice to yield the smallest statistical error bound for the eigenspace estimation. In particular, we can easily show the three terms $C_{q,q'}$, $\xi_{q,q'}$ and $(q+q'-1)\xi_{q,q'}$ involving $q,q'$ all attain their minimum at  $q=q'=\frac{\mathrm{e}^\epsilon}{1+\mathrm{e}^\epsilon}$. For example, consider $\xi_{q,q'}$, plugging in the upper boundaries shown in Figure \ref{region}, i.e., $q'=1-\frac{q}{\mathrm{e}^\epsilon}$ when $q\leq \frac{\mathrm{e}^\epsilon}{1+\mathrm{e}^\epsilon}$ and $q'=\mathrm{e}^\epsilon (1-q)$ when $q\geq \frac{\mathrm{e}^\epsilon}{1+\mathrm{e}^\epsilon}$, we have
\[
\xi_{q,q'}=\frac{1-q'}{(q+q'-1)^2} = \left\{
\begin{array}{ll}
\frac{1}{q\mathrm{e}^\epsilon(1-\mathrm{e}^{-\epsilon})^2} & \text{if } q \leq \frac{\mathrm{e}^\epsilon}{1+\mathrm{e}^\epsilon},\\
\frac{1-\mathrm{e}^\epsilon +\mathrm{e}^\epsilon q}{(1-q)^2 (1-\mathrm{e}^\epsilon )^2} & \text{if } q \geq \frac{\mathrm{e}^\epsilon}{1+\mathrm{e}^\epsilon},
\end{array}
\right.
\]
which attains minimum at the $q=\frac{\mathrm{e}^\epsilon}{1+\mathrm{e}^\epsilon}$ and at this time $q'$ equals $q$. The terms $C_{q,q'}$ and $(q+q'-1)\xi_{q,q'}$ can be derived similarly.
\end{remark}

The next corollary shows that under certain extra conditions, the bound of our distributed estimator reduces to its centralized counterpart.

\begin{corollary}
\label{corofcen}
Suppose the assumptions and conditions in Theorem \ref{sparsecor} all hold. If we further have
\begin{equation}
\label{mrequire}
m\lesssim  \frac{n\rho^2}{(C_{q,q'}\rho+\xi_{q,q'})\log (s+n)}\cdot L^{1/2};
\end{equation}
\begin{equation}
\label{hsparse}
\mathcal H\lesssim \max \bigg\{\frac{n^{1/2}\rho^2 (C_{q,q'}\rho+ \xi_{q,q'})^{1/2}\log^{1/2}(L+n)}{L^{1/4}\rho},\, {n^{1/2}\rho }(\rho+(q+q'-1)\xi_{q,q'})\bigg\},
\end{equation}
then $\mathcal E_3$ dominates $\mathcal E_1$ and $\mathcal E_2$; see their definitions in Theorem \ref{sparsecor}.
The output of $\tilde{V}$ of Algorithm \ref{Oneshotsvd} satisfies
\begin{equation*}
\mathrm{dist} (\tilde{V}, V)\lesssim \max \bigg\{\frac{\log(L+n)(C_{q,q'}\rho+\xi_{q,q'})}{n{\sqrt{L}}\rho^2}, \;\frac{(1+(q+q'-1)\xi_{q,q'}/\rho)^2}{n}
\bigg\}
\end{equation*}
with probability larger than $1-O((L+n)^{-\nu})-O(m(s+n)^{-\nu}))$ for some constant $\nu>0$.
\end{corollary}

We make the following remarks on Corollary \ref{corofcen}.
\begin{remark}
Condition (\ref{mrequire}) requires the number of machines should not be large given other parameters. In particular, when ${n}\rho^2/(C_{q,q'}\rho+\xi_{q,q'})\asymp {{\log(s+n)}}$, (\ref{mrequire}) reduces to $m\lesssim \sqrt{L}$, which is a crucial condition for the distributed estimator to match the error rate of its centralized counterpart in one-shot-based distributed learning literature, see \citet{zhang13communication,lee2017communication}, among others. The requirement for $m$ might be alleviated if one considers multi-round communications; see \citet{jordan2018communication,chen2020distributed} in the context of regression.

Condition (\ref{hsparse}) is mild. For example, suppose $\xi_{q,q'}\lesssim C_{q,q'}\rho$ and $\rho \asymp \frac{\log (s+n)m}{n\sqrt{L}}$ (i.e., the boundary in (\ref{mrequire})), then (\ref{hsparse}) requires $\mathcal H\lesssim  \frac{m^{2}\log^{1/2}(L+n)\log ^{3/2}(s+n)}{nL}$. On the other hand, we naturally have $\mathcal H\lesssim n\rho^2\asymp \frac{m^2\log^2(s+n)}{nL}$ (see the proof of Theorem \ref{sparsecor}).
\end{remark}

\begin{remark}
The error bound is comparable to  \citet{lei2020bias} when $\xi_{q,q'}\lesssim C_{q,q'}\rho$. In particular, if $\rho \asymp \frac{\log (s+n)m}{n\sqrt{L}}$ and $\epsilon\asymp \log(\frac{n\sqrt{L}}{m})$ (suppose $q,q'$ relates to $\epsilon$ via (\ref{eq})), then $\mathcal E_3$ reduces to $\max\{\frac{\log\, (L+n)}{n\rho\sqrt{L}},\, \frac{1}{n} \}$, which is identical (up to log terms) to that in \citet{lei2020bias}. Here, we remark that the large $\epsilon$ in the context of local-DP \citep{duchi2018minimax} (i.e., DP at the level of original data rather than aggregated data) is reasonable and has also been used in other literature.
We note that \citet{seif2022differentially} chose $\epsilon \gtrsim \log\, n$ when they studied the strong consistency of community detection algorithms under RR-perturbed SBMs. We also note that the work by \citet{bhowmick2018protection} revisited the protections of local-DP \citep{duchi2018minimax} and it is said that algorithms that only protect against the accurate reconstruction of (functions of) an individual's data could allow a {\emph{larger}} privacy budget ($\epsilon\gg 1$).
\end{remark}

The following theorem shows that the misclassification error bound for \texttt{ppDSC} turns out to be the squared error of the eigen-vectors $\tilde{V}$.

\begin{theorem}
\label{localmis}
Suppose the conditions in Corollary \ref{corofcen} all hold. Denote the set of misclassified nodes by $\mathcal M$.
Then, with probability larger $1-O((L+n)^{-\nu})-O(m(s+n)^{-\nu}))$ for some constant $\nu>0$, Algorithm \ref{Oneshotsvd} has misclassification rate upper bounded as
$\frac{|\mathcal M|}{n}\lesssim\mathcal E_3^2;$
see the definition of $\mathcal E_3$ in Theorem \ref{sparsecor}.
\end{theorem}

Finally, we discuss the necessity of the bias-adjustment procedure in Section \ref{app::theory} of the Supplementary Materials. We investigate the deviation of the non-debiased matrix $\tilde{M}:=\frac{1}{L}\sum_l\frac{1}{n} \tilde{A}_l^2$ from population matrices $Q$ and $\tilde{Q}:=\frac{1}{L}\sum_l\left((q+q'-1) P_l+(1-q')\textbf{1}_n \textbf{1}_n^\T\right)^2/{n}$, respectively.
It turns out that the bias-adjustment procedure can lead to a smaller eigenvector and misclassification error bound, no matter which population matrix is used.

\section{Simulations}
\label{sec:sim}
In this section, we evaluate the finite sample performance of the proposed algorithm. We show how the number of networks, the number of nodes, and the number of local machines affect the finite sample behavior of the resulting estimates. We also justify
our theory in terms of how to choose the edge-flipping probabilities in RR under a given privacy budget and how the network sparsity affects the performance under different privacy budgets.
In Section \ref{app::exp} of the Supplementary Materials, we test the effect of heterogeneity and network sparsity on the performance of the proposed algorithm.

\paragraph{Practical consideration.} To mimic the scenario considered in this work, we assume that all the generated network datasets are private and locally distributed. To preserve privacy, we assume that the data analyst can not use the original data directly but can only use the \emph{RR-perturbed} networks. For the local computation, we generally adopt the distributed SVD framework \citep{charisopoulos2021communication} considered in this work.

\paragraph{Methods compared.} \textcolor{black}{Our algorithm is denoted by \texttt{ppDSC}. We compare our algorithm with the following two \emph{distributed} algorithms denoted by \texttt{ppDSC-1b} and \texttt{ppDSC-2b}, both of which follow the above practical restrictions. Hence, \texttt{ppDSC-1b} and \texttt{ppDSC-2b} suffer from the model misspecification problem. We also compare two baseline \emph{centralized} algorithms. One is the centralized counterpart of the proposed algorithm, denoted by \texttt{ppSC}. The other is the algorithm in \citet{lei2020bias}, denoted by \texttt{Oracle} because it knows the unperturbed network like an oracle. }
\begin{itemize}
    \item \texttt{ppDSC-1b}: \textcolor{black}{computed according to Algorithm \ref{Oneshotsvd} but with line \ref{debiasline} replaced by one-step bias-correction (i.e., diagonal-debiasing) given in \citet{lei2020bias} (i.e., $\tilde{A}_l^2-\mbox {diag}(\tilde{A}_l^2)$);}
    \item \texttt{ppDSC-2b}: computed according to Algorithm \ref{Oneshotsvd} but without any bias-correction, which corresponds to \citet{charisopoulos2021communication} with RR-perturbed networks as input;
    \item \texttt{ppSC}: obtained from the spectral clustering based on the \emph{debiased} sum of squared RR-perturbed matrices  (i.e., $\overline{M}$ defined in (\ref{barM})).
    \item \texttt{Oracle}: \textcolor{black}{the algorithm in \cite{lei2020bias}; obtained from the spectral clustering based on the diagonal-debiased sum of squared \emph{original} adjacency matrices (i.e., $\sum_{l}(A_l^2-{\rm diag}(A_l^2)))$.}
\end{itemize}

\paragraph{Performance measures.} According to our theoretical analysis, we use the \emph{projection distance} to measure the subspace distance between the estimated eigenvectors and the true eigenvectors and use the \emph{misclassification rate} to measure the proportion of misclassified (up to permutations) nodes.

\paragraph{Network-generation.} The local networks are generated independently from multi-layer SBMs with the structure given in Section \ref{sec::alg}. Similar to \citet{lei2020bias}, we fix $K=3$ with balanced communities and set $B_l:=0.8B^{(1)}$ for $l=1,...,\lfloor L/2\rfloor$ and $B_l:=0.6B^{(2)}$ for $l=\lfloor L/2\rfloor+1,...,L$ with

{\begin{equation*}
\scriptsize
B^{(1)}=W\begin{bmatrix}
1.5 &&   \\
 & 0.2&\\
 &&0.4
\end{bmatrix}
W^\T\approx
\begin{bmatrix}
0.62&0.22&0.46\\
0.22&0.62&0.46\\
0.46&0.46&0.85
\end{bmatrix};
\end{equation*}}
 \begin{equation*}
 \scriptsize
B^{(2)}=W\begin{bmatrix}
1.5 &&   \\
 & 0.2&\\
 &&-0.4
\end{bmatrix}
W^\T\approx
\begin{bmatrix}
0.22&0.62&0.46\\
0.62&0.22&0.46\\
0.46&0.46&0.85
\end{bmatrix},
\end{equation*}
where
\begin{equation*}
\scriptsize
W=
\begin{bmatrix}
1/2&1/2&-\sqrt{2}/2\\
1/2&1/2&\sqrt{2}/2\\
\sqrt{2}/2&-\sqrt{2}/2&0
\end{bmatrix}.
\end{equation*}

\paragraph{Experiment 1.} We evaluate the effect of $L$, $n$ and $m$ on the numerical performance. The parameter setup is as follows.
\begin{itemize}
\item  Effect of $L$: $n=210$, $q=q'=0.8$, $m=L$, and $L$ from 2 to 17.
\item  Effect of $n$: $L=m=12$, $q=q'=0.8$, and $n$ ranges from 60 to 230.
\item  Effect of $m$: $L=12$, $n=210$, $q=q'=0.8$, and $m$ varies in $\{2,4,6,12\}$.
\end{itemize}

The averaged results over 20 replications are displayed in Figure \ref{algvaluation}. We have the following observations. First, as expected, as $L,n$ increases and $m$ decreases, the performance of all estimators improves. Second, our estimator \texttt{ppDSC} is superior to the biased \texttt{ppDSC-1b} and \texttt{ppDSC-2b}. It is because \texttt{ppDSC-1b} and \texttt{ppDSC-2b} ignore the bias caused by privacy-preserving the networks and thus mis-specify the model. Third, as with extra local computation and privacy preservation demand, our method \texttt{ppDSC} is generally not better than the centralized methods \texttt{ppSC} and \texttt{Oracle}. However, when $L$ and $n$ become large, and $m$ becomes small, they are getting close. This is consistent with our theoretical results.

\begin{figure}[h]{}
\centering
\centerline{\small(I) Projection distance }
\subfigure[Effect of $L$]{\includegraphics[height=4.5cm,width=4.5cm,angle=0]{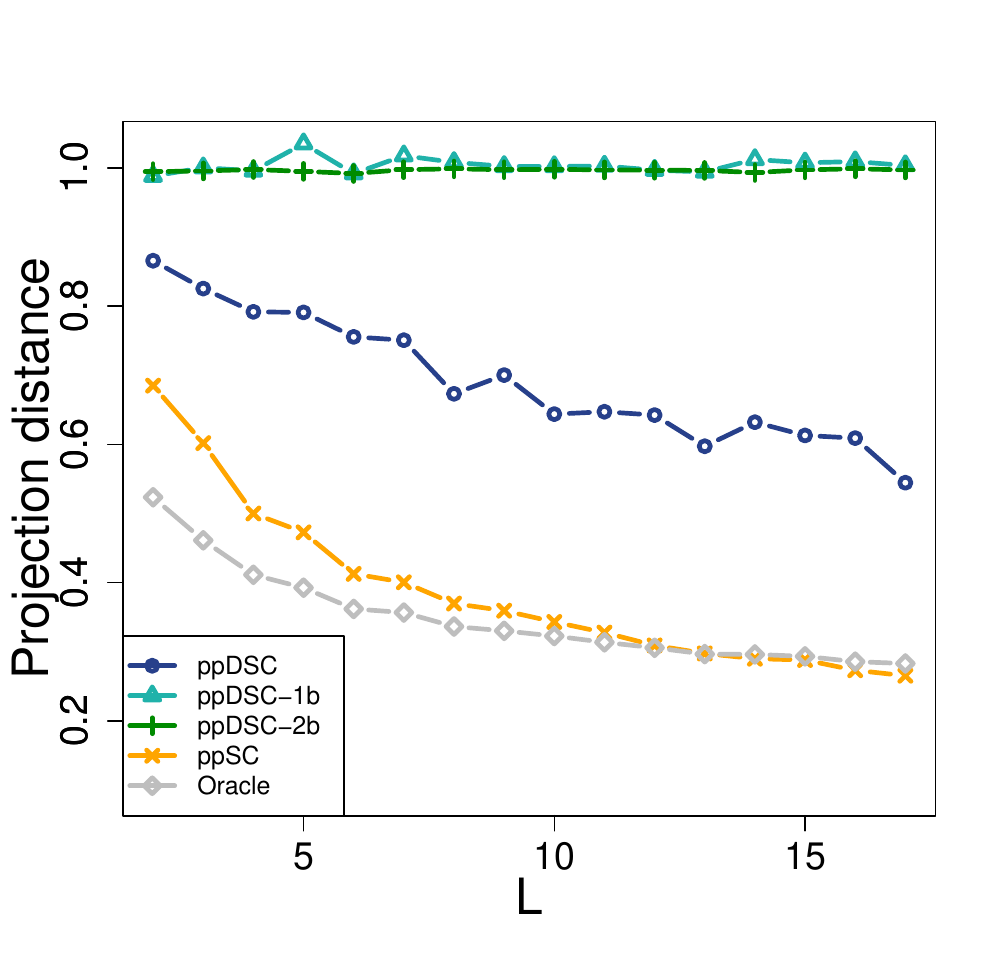}}
\subfigure[Effect of $n$]{\includegraphics[height=4.5cm,width=4.5cm,angle=0]{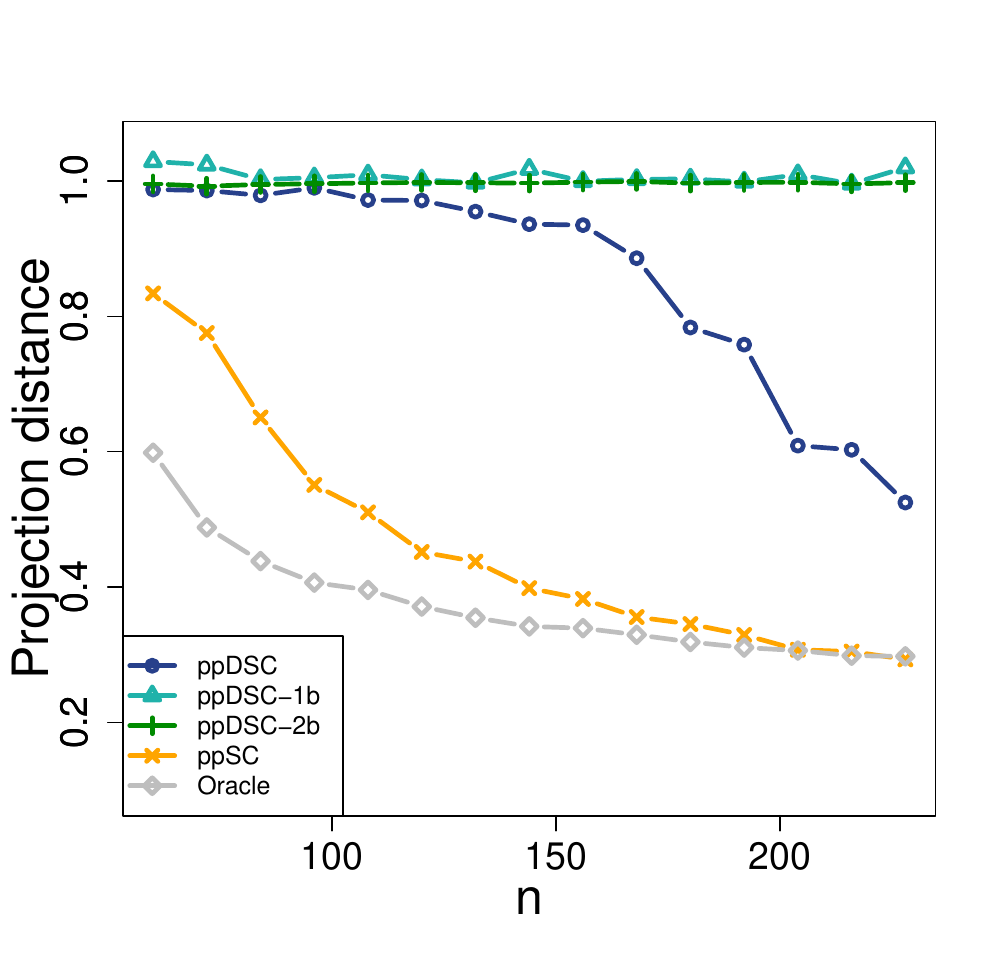}}
\subfigure[Effect of $m$]{\includegraphics[height=4.5cm,width=4.5cm,angle=0]{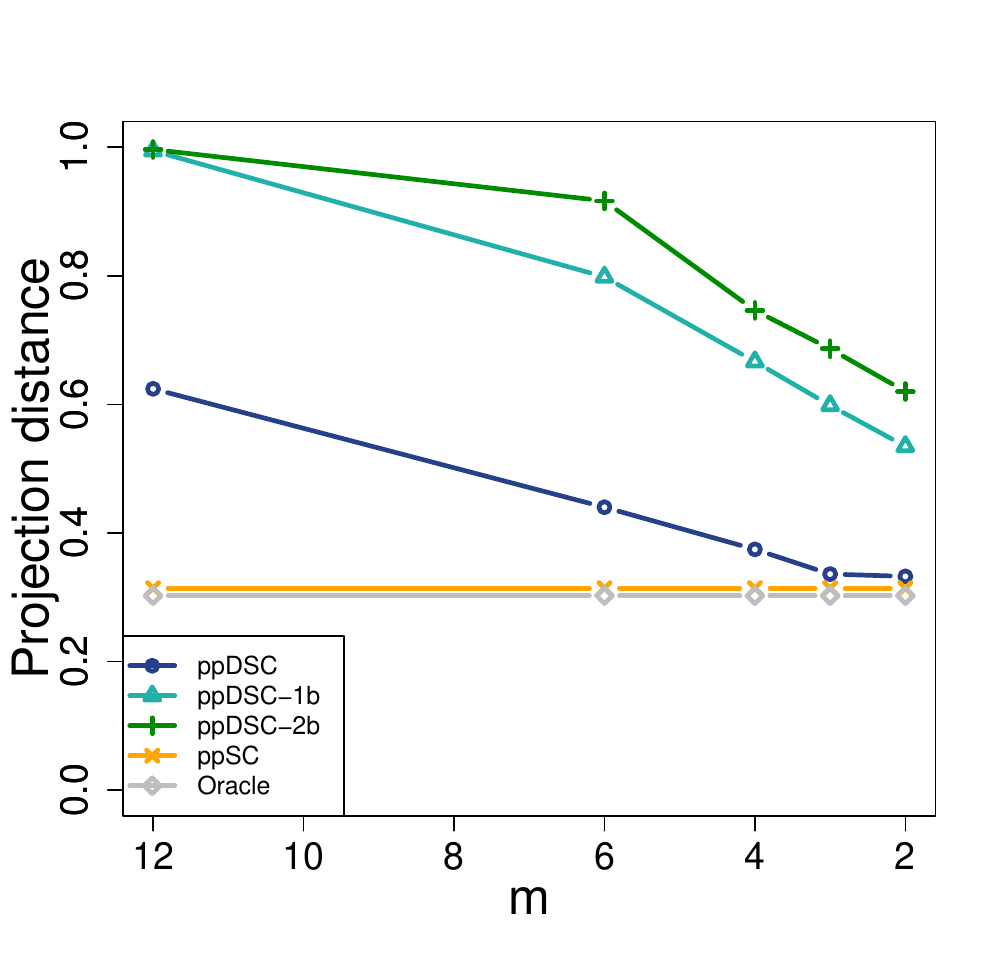}}
\vspace{0.5cm}\\
\centerline{\small(II) Misclassification rate }
\subfigure[Effect of $L$]{\includegraphics[height=4.5cm,width=4.5cm,angle=0]{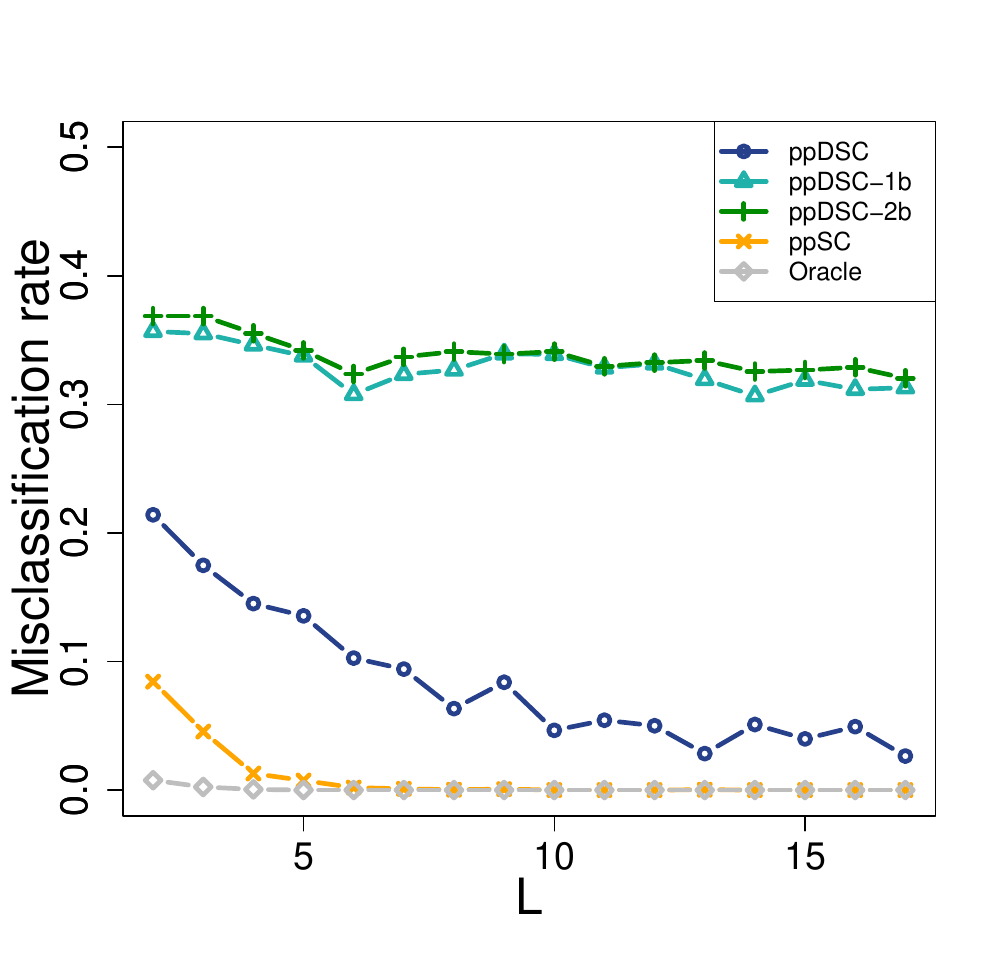}}
\subfigure[Effect of $n$]{\includegraphics[height=4.5cm,width=4.5cm,angle=0]{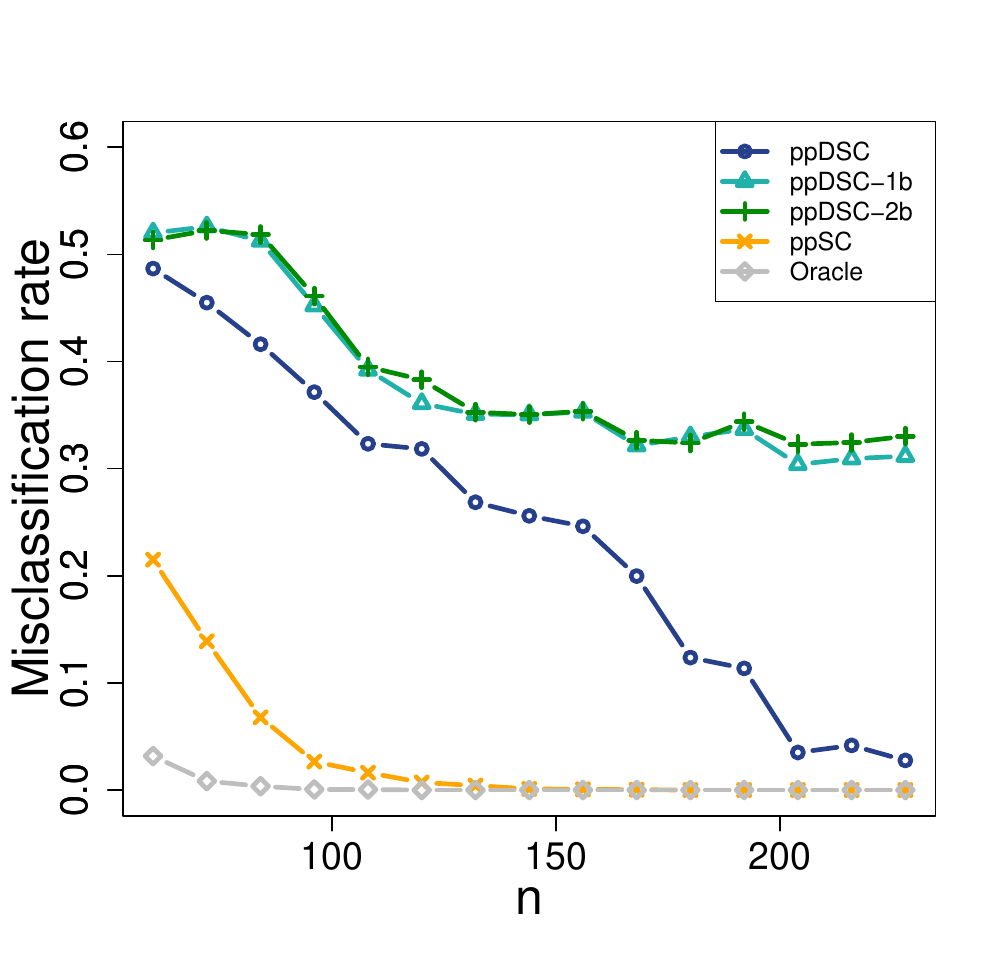}}
\subfigure[Effect of $m$]{\includegraphics[height=4.5cm,width=4.5cm,angle=0]{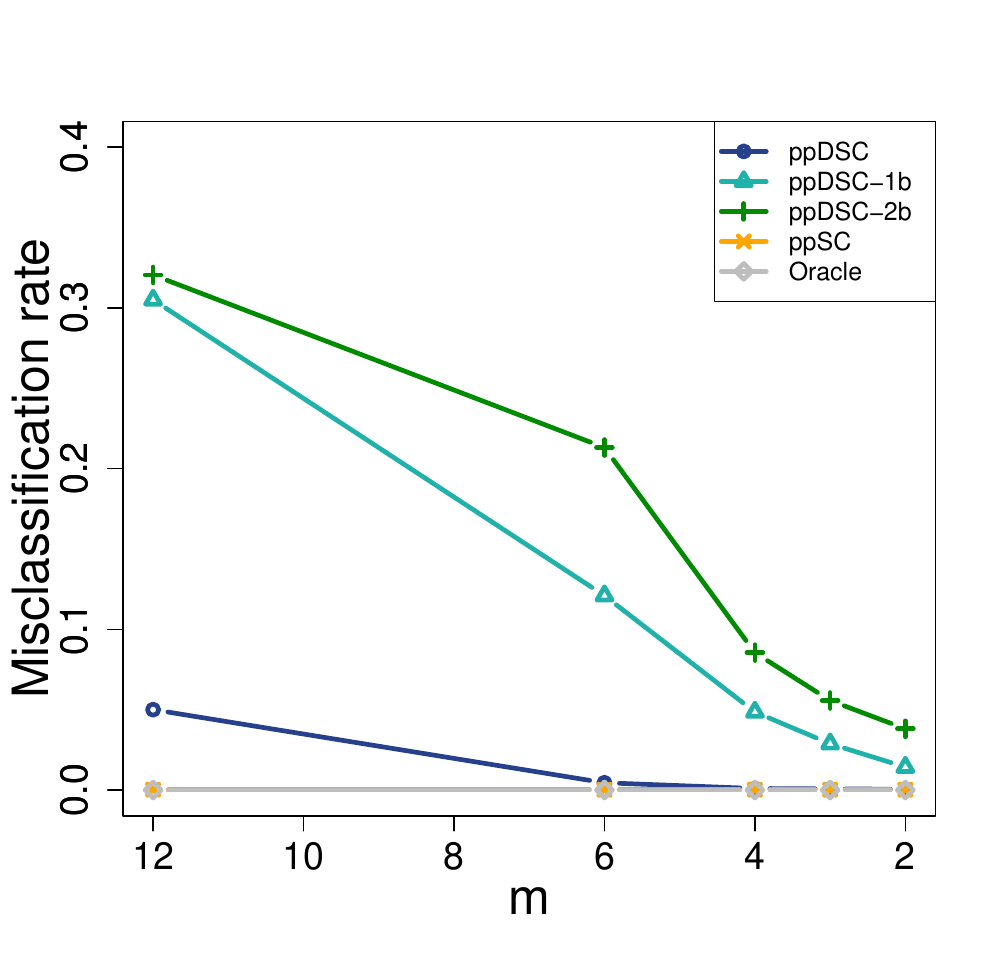}}
\caption{Comparison of \texttt{ppDSC}, \texttt{ppDSC-1b}, \texttt{ppDSC-2b}, \texttt{ppSC} and \texttt{Oracle} on the simulated data. The effect of the number of networks $L$, the number of nodes $n$, and the number of local machines $m$ are shown, respectively. The projection distance and misclassification rate are evaluated. }\label{algvaluation}
\end{figure}

\paragraph{Experiment 2.} \textcolor{black}{We evaluate the performance of \texttt{ppDSC} with the privacy budget $\epsilon=1$ fixed and the edge-fillping probabilities $q,q'$ varying in the feasible region of DP shown in Figure \ref{region}. Other parameters are $n=210$ and $L=m=12$. The averaged misclassification rates over 10 replications under different couples of $q$ and $q'$ are displayed in Figure \ref{bestqvaluation}, where the infeasible region is filled by pure white. When $q,q'$ are getting close to the upper boundaries, the performance of \texttt{ppDSC} becomes better. In particular, the misclassification rate attains the empirical minimum at the position marked by the pink cross, which is quite close to the theoretically best choice of $q=q'=\frac{{\rm e}^\epsilon}{1+{\rm e}^\epsilon}$ (see the discussions after Theorem \ref{centralbound}) marked by the red star. }

\begin{figure}[h]{}
\centering
{\includegraphics[height=5.5cm,width=7.4cm,angle=0]{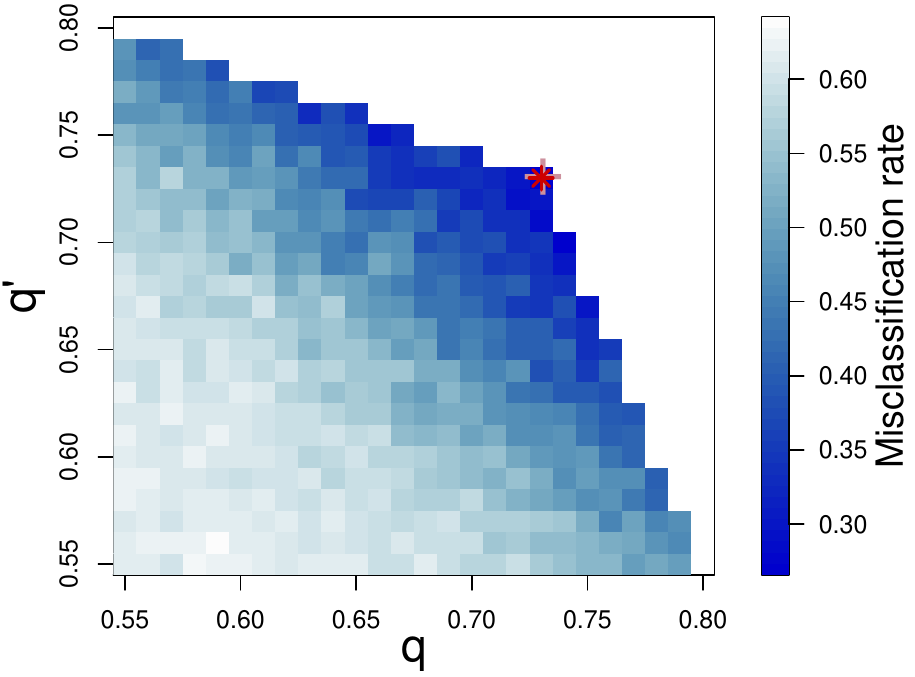}}
\caption{The clustering performance of \texttt{ppDSC} with a pre-specified $\epsilon=1$ and $q,q'$ varying in the feasible region of DP (see Figure \ref{region}). The empirically and theoretically best combination of $q,q'$ are marked with red star and pink cross, respectively. }\label{bestqvaluation}
\end{figure}

\section{Real data analysis}
\label{sec::real}
In this section, we analyze the \texttt{AUCS} dataset \citep{rossi2015towards}, which is a public dataset collected by the employees at the Department of Computer Science of Aarhus University, to demonstrate the utility of the proposed algorithm. Though privacy preservation is prominent nowadays, to the best of our knowledge, networks known to be noisy and perturbed by the RR mechanism are not currently available to the public. Hence, we regard the \texttt{AUCS} dataset as the original data without any perturbation. We will mimic the actual setup to see how the privacy and local computation affect the clustering performance of the proposed algorithm.

\vspace{-0.4cm}
\paragraph{Data description.} There are 61 employees (out of the total number of 142) who decided to join the survey, consisting of professors, postdoctoral researchers, Ph.D. students, and administration staff. Five types of relationships among the employees are measured. Namely, current working relationships, repeated leisure activities, regularly eating lunch together, co-authorship of a publication, and friendship on Facebook. The first three relationships were collected via a questionnaire. All the respondents answered all questions, indicating that the data was complete. And the last two relationships were acquired online via the DBLP database and Facebook. As a result, the original dataset consists of 5 undirected networks with 61 common nodes.
Besides the relationships, the research group labels that 55 out of 61 employees are available. There are eight research groups ($K=8$) in total. Our analysis uses the subnetworks of 55 employees with known labels.
\vspace{-0.4cm}
\paragraph{Evidence of communities and heterogeneity.} The topology of networks with respect to different relationships is shown in Figure \ref{localnetwork} in the Supplementary
Materials with colors indicating the research groups. It is clear to see that the network structures are quite different. To take a closer look at the community structures, we reorder the nodes according to their underlying research groups (i.e., nodes corresponding to the same research group are aligned next to each other). We further plot the corresponding network adjacency matrices (see Figure \ref{localmatrix} in the Supplementary
Materials), where the whole network (Figure \ref{localmatrix}(a)) corresponds to the network that contains the edges of all five relationships. The whole network shows conspicuous block structures in that nodes in the same research group have more links than those across different research groups. Hence, it is reliable to use the research group labels as the underlying community memberships. Distinct from the whole network, we see that the community structures of each network layer are heterogeneous in that some communities are not evident or even missing to different extents.


\vspace{-0.4cm}
\paragraph{Practical consideration.} Though the \texttt{AUCS} dataset is publicly available, we note that some information about individuals say the eating and working relationships are sometimes sensitive and private. In addition, different relationships may be investigated and owned by different parties. Therefore, we consider the setting that the networks corresponding to each kind of relationship are private and locally owned by different parties. We will see how the privacy and local computation constraints affect the clustering performance of the proposed algorithm.
\vspace{-0.4cm}
\paragraph{Methods compared.} Consistent with the simulation experiments, we compare our method \texttt{ppDSC} with \texttt{ppDSC-1b}, \texttt{ppDSC-2b}, \texttt{ppSC} and \texttt{Oracle}. The detailed description of each method can be found in Section \ref{sec:sim}. We study how the performance of each method alters with the network number $L$ and the privacy parameters $q$ and $q'$. Specifically, we fix $q(q')$ to be 0.95 and vary $q'(q)$ to verify the different effect of $q$ and $q'$ indicated by the theory. To save space, the detailed descriptions on the experimental set-up is relegated to Section \ref{app::exp} of the Supplementary Materials.

\vspace{-0.4cm}
\paragraph{Results.} The averaged misclassification rates over 20 replications are displayed in Figure \ref{aucsplot}. We have the following observations. First, as $L$ increases, the misclassification rate of all five methods decreases, showing the benefits of aggregating the information of local networks. Note that when $L=5$, a small drop in terms of the clustering performance appears, partially because the fifth network (i.e., the coauthor network) is rather sparse and has a weak signal of communities. Second, as expected, as $q$ and $q'$ increase, the clustering performance of all four privacy-preserving methods becomes better. We note that the clustering performance with $q'$ fixed to be 0.95 (See Figure \ref{aucsplot}(c)) is better than that with $q$ fixed to be 0.95 (See Figure \ref{aucsplot}(d)), which shows that large $q'$ is more desirable than large $q$. In other words, perturbing 0 to 1 is more harmful than perturbing 1 to 0. This is consistent with the theoretical results. Third, consistent with the simulation results, the proposed method \texttt{ppDSC} is better than the biased method \texttt{ppDSC-1b} and \texttt{ppDSC-2b} and is close to the baseline methods \texttt{ppSC} and \texttt{Oracle} as $q$ and $q'$ increase.


\begin{figure*}[!h]{}
\centering
\subfigure[Effect of $L$]{\includegraphics[height=3.5cm,width=3.6cm,angle=0]{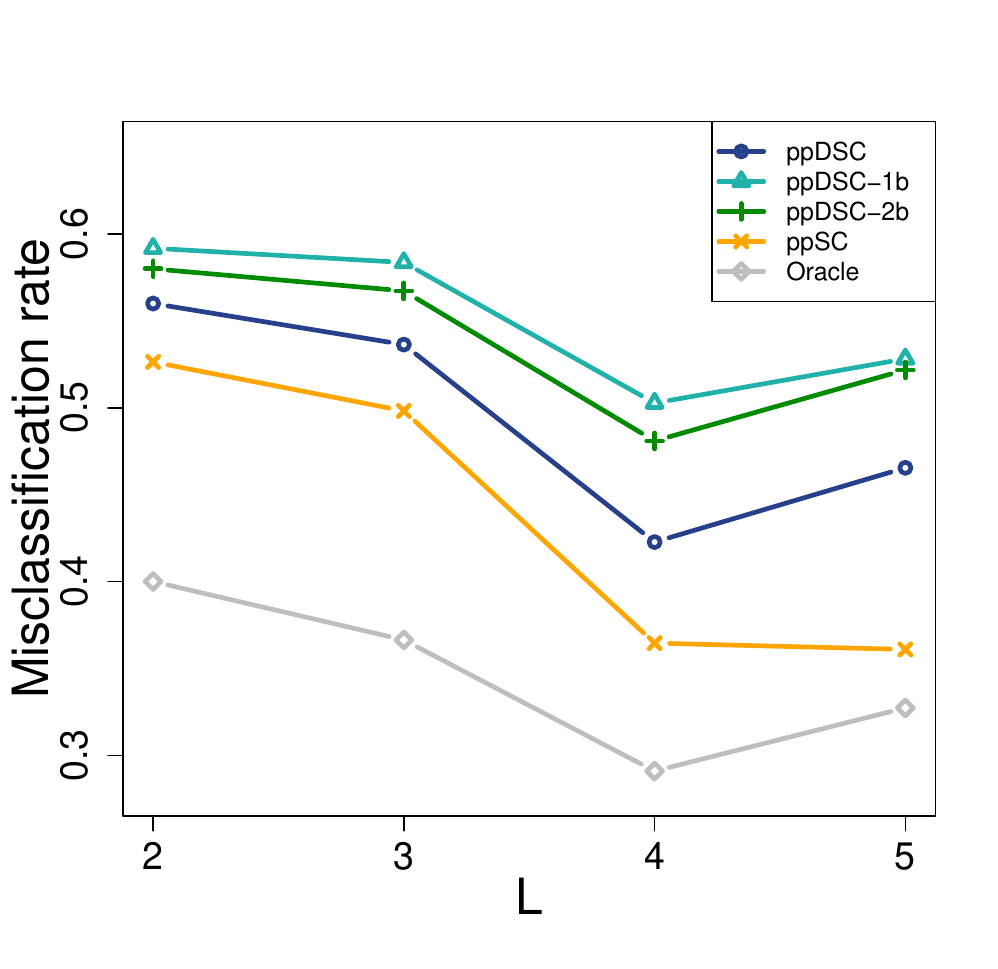}}
\subfigure[Effect of $q,q'$]{\includegraphics[height=3.5cm,width=3.6cm,angle=0]{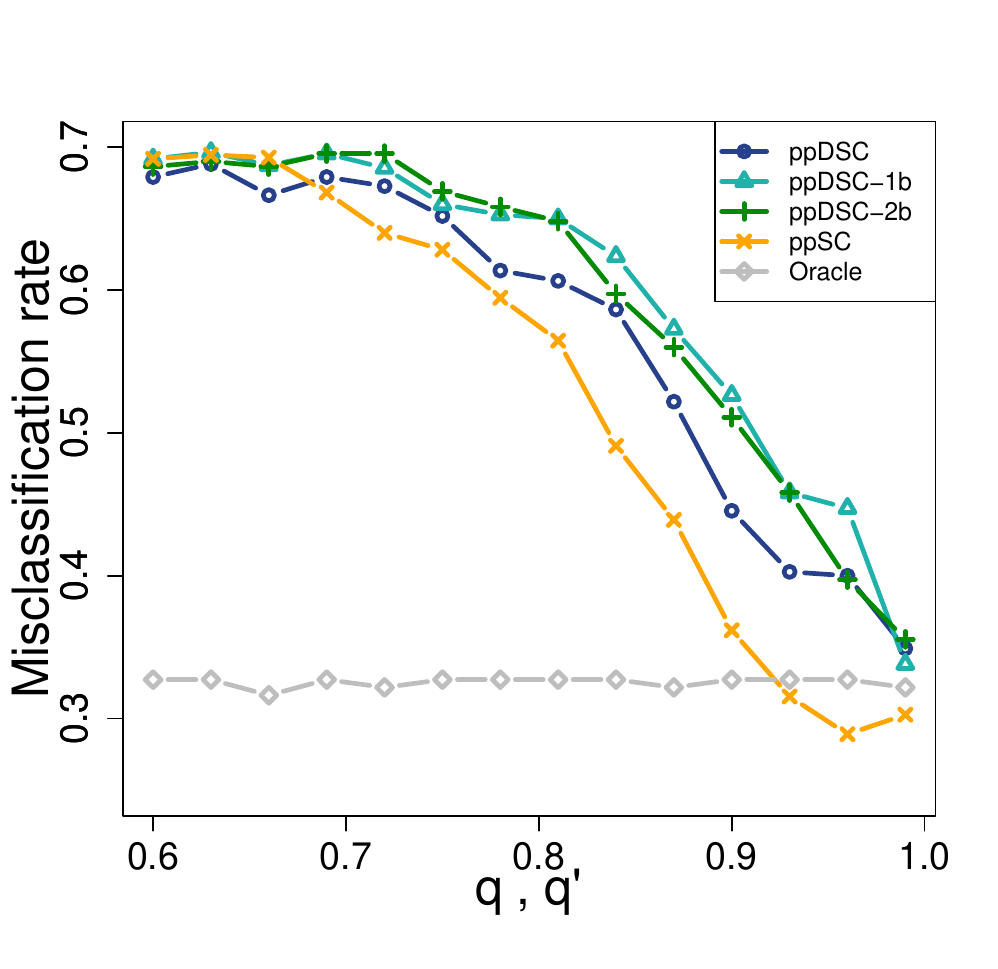}}
\subfigure[Effect of $q$]{\includegraphics[height=3.5cm,width=3.6cm,angle=0]{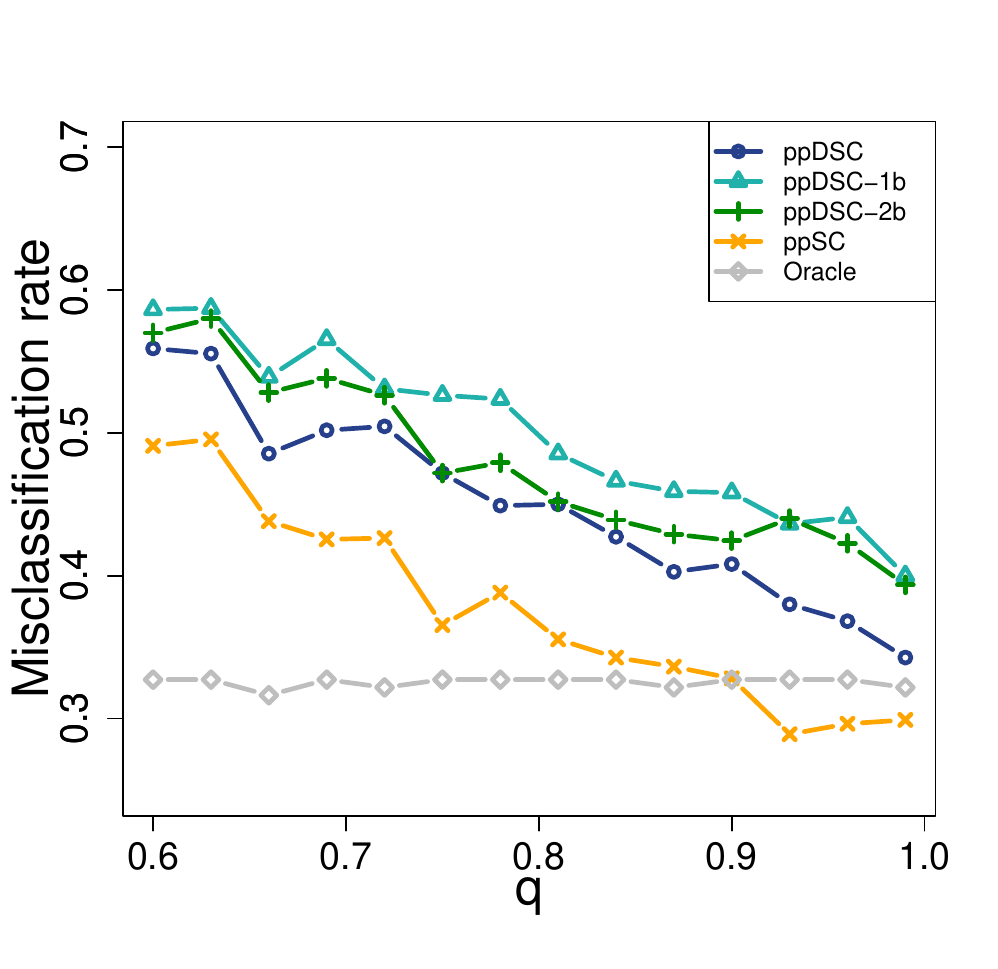}}
\subfigure[Effect of $q'$]{\includegraphics[height=3.5cm,width=3.6cm,angle=0]{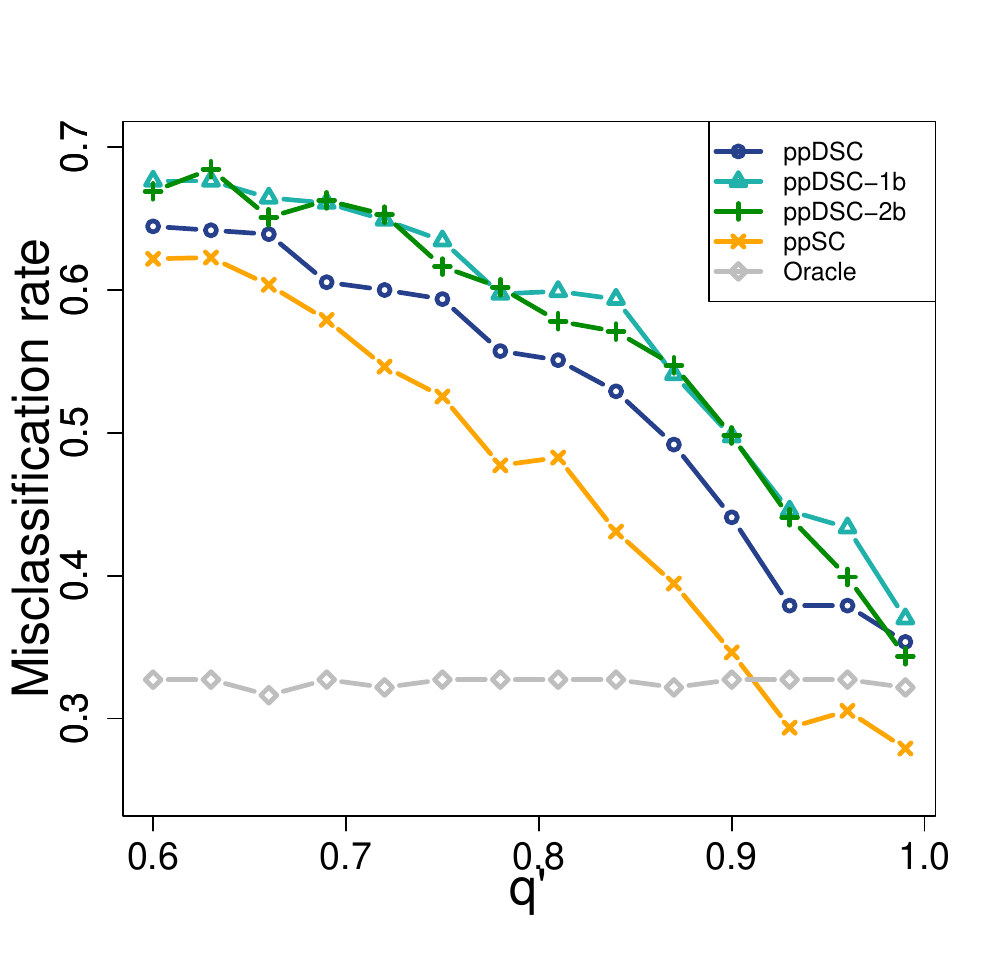}}
\caption{Comparison of \texttt{ppDSC}, \texttt{ppDSC-1b}, \texttt{ppDSC-2b} and \texttt{Oracle} on the \texttt{AUCS} network. The effect of the privacy parameters $q,q'$ ($q,q'$ synchronously vary, $q'$ fixed but $q$ varies, $q$ fixed but $q'$ varies), and the number of networks $L$ are shown, respectively. The misclassification rate with respect to the research group labels is evaluated.}\label{aucsplot}
\end{figure*}

\section{Conclusion}
\label{sec::concl}

Driven by practical data collection and the associated analytic challenges, we provide a powerful toolbox for community detection and estimation in multi-layer SBMs based on locally stored networks with privacy protection. In the toolbox, a new algorithm named  \texttt{ppDSC} has been developed, and it is shown to be both communication-efficient and privacy-aware. We also propose a two-step bias-adjustment procedure to remove the biases caused by the RR perturbation and taking the square of the debiased adjacency matrices. The procedure is found to be effective both theoretically and numerically. Moreover, we have rigorously investigated the misclassification rate of the algorithm. In particular, we show that when the number of machines and the effective heterogeneity are well-controlled, the misclassification rate of our distributed algorithm matches its centralized counterpart. More interestingly, our result indicates the best choice of privacy parameters under a given privacy budget $\epsilon$.
In addition, we demonstrate the effect of heterogeneity among population networks and the asymmetrical roles of the privacy parameters $q$ and $q'$ on the estimation accuracy.
The numerical experiments support our theoretical findings well.

The proposed methodology can be extended to a variety of scenarios. First, the \texttt{ppDSC} algorithm involves multiple steps. The methods used in some steps are quite general and can be applied in other contexts. For example, the proposed bias-adjustment procedure can be used in other stochastic block models in different distributed SVD frameworks, e.g., \citet{fan2019distributed,arroyo2021inference}. The method for computing the local eigenspaces can also be replaced with the tensor-based method, e.g., \citet{jing2021community}. Second, it is well-known that privacy could be amplified by sampling \citep{balle2018privacy}. With the same level of noise from privacy preservation, a DP algorithm with sampled data leaks less privacy than that with all data. Therefore, it is desirable to study the trade-off between privacy gain and information loss when sampling the network entries in network analysis. Third, for statistical convenience, we consider the local-DP \citep{duchi2018minimax} protection of networks. We may extend the proposed methods in the context of central DP. Fourth, we focus on the distributed learning framework with only one round of communication. It is desirable to extend the results to the distributed framework with multiple rounds, say the communication-efficient distributed power method \citep{guo2021privacy}.
Last but not least, the method and theory for estimating $K$ can be an interesting topic to explore in the future. \citep{fishkind2013consistent,li2020network,ma2021determining}.

\section*{Appendix}

\appendix

\numberwithin{figure}{section}
\numberwithin{equation}{section}
\numberwithin{theorem}{section}
\phantomsection

\spacingset{1.75}

Section \ref{app::biasadj} provides the details of the bias-adjustment procedure.
Section \ref{app::theory} includes the additional theoretical results and discussions.
Section \ref{app::exp} presents the additional simulation experiments and the results of real data analysis.
Section \ref{app::proof} contains the technical proofs. Several auxiliary lemmas are provided in Section \ref{app::lemma}.

\section{Details of bias-adjustment}
\phantomsection

\label{app::biasadj}
We illustrate how the diagonal elements of $\overline{A}_l^2$ would cause bias for $P_l^2$, even though $\overline{A}_l$ is the unbiased estimator of $P_l$.
Denote the non-diagonal part of $P_l$ by $P_{l1}:=P_l-{{\diag}}(P_l)$, and denote the deviation of $\overline{A}_l$ from $P_{l1}$ by $X_l:=\overline{A}_l-P_{l1}$. After simple calculation, we obtain
{\small
\begin{align}
\label{composion1}
\overline{A}_l^2&=(X_l+P_{l1})^2=X_l^2+P_{l1}^2+X_l P_{l1}+P_{l1} X_l.
\end{align}
}
Plugging in
$P_{l1}^2=(P_l-{\diag}(P_l))^2$,
we have
{
\begin{align}
\label{composion2}
\overline{A}_l^2-P_l^2&=E_{l,0}+E_{l,1}+E_{l,2}+E_{l,3}
\end{align}
}
with
\begin{align}
E_{l,0}&:={-P_l\diag(P_l)-\diag(P_l)\cdot P_l +(\diag(P_l))^2},\nonumber\\
E_{l,1}&:=X_l(P_l-{\diag}(P_l))+ (P_l-\diag(P_l))X_l,\nonumber\\
E_{l,2}&:=X_l^2-{\diag}(X_l^2),\nonumber\\
E_{l,3}&:={\diag}(X_l^2).\nonumber
\end{align}
We would see in Appendix \ref{app::proof} that the first three terms are actually all small. While for the fourth term $E_{l,3}$, we now proceed to bound it.
Note that
\begin{align}
\label{e3bound}
(E_{l,3})_{ii}&=\sum_{j=1}^n X_{l,ij}^2=\sum_{j=1,j\neq i}^n (\overline{A}_{l,ij}-P_{l,ij})^2\nonumber\\
&=\frac{1}{(q+q'-1)^2}\sum_{j=1}^n \mathbb I(\tA_{l,ij}=1)(1-(1-q')-(q+q'-1)P_{l,ij})^2\nonumber\\
&\quad\quad\quad\quad\quad\quad+\frac{1}{(q+q'-1)^2}\sum_{j=1}^n \mathbb I(\tA_{l,ij}=0)(1-q'+(q+q'-1)P_{l,ij})^2\nonumber\\
&\leq  \frac{\sum_{j=1}^n \mathbb I(\tA_{l,ij}=1)(q')^2}{(q+q'-1)^2}+\frac{n(1-q'+(q+q'-1)\kappa)^2}{(q+q'-1)^2}\nonumber\\
&= \frac{\sum_{j}\tA_{l,ij}(q')^2}{(q+q'-1)^2}+\frac{n(1-q'+(q+q'-1)\kappa)^2}{(q+q'-1)^2},
\end{align}
where in the first inequality we assume that $q,q'>1/2$ and $\kappa=\max_{i,j,l}P_{l,ij}$. Hence, the error would be smaller if we further subtract $\frac{\sum_{j}\tA_{l,ij}(q')^2}{(q+q'-1)^2}$ from the $i$th diagonal elements of $\overline{A}_l^2$ and assume $\kappa=o(1)$. More importantly, as we aim to approximate $\sum_{l=1}^L P_l^2$, this debiasing step is crucial when $L$ is increasing.

\section{The effect of bias-adjustment}
\phantomsection
\label{app::theory}
We theoretically show the effect of bias-adjustment. Denote the non-debiased estimator by $$\tilde{M}:=\frac{1}{L}\sum_l\frac{1}{n} \tilde{A}_l^2$$ and the corresponding
population matrix by $$\tilde{Q}:=\frac{1}{L}\sum_l\left((q+q'-1) P_l+(1-q')\textbf{1}_n \textbf{1}_n^\T\right)^2/{n}.$$ The following proposition provides the estimation error of $\tilde{M}$.

\begin{proposition}
\label{nondebias-theory}
Suppose Assumption \ref{sparse} and condition (\ref{eandl}) holds. Define
\begin{align*}
\small\mathcal I:=\max \Big\{\frac{(q\rho+1-q')^{1/2}n^{1/2}[\rho+(1-q')/(q+q'-1)] \log^{1/2}(L+n)}{(q+q'-1)\sqrt{L}},\nonumber\\
\quad\quad\quad\quad\quad\quad\quad\quad\frac{(q\rho+1-q')\log(L+n)}{(q+q'-1)^2\sqrt{L}},\, \frac{((q+q'-1)\rho+1-q')^2}{(q+q'-1)^2}\Big\},
\end{align*}
then the non-debiased estimator $\tilde{M}$ of $\tilde {Q}$ satisfies
\begin{equation}
\phantomsection
\label{nondebias}
\|\tilde{M}-\tilde{Q}\|_2\lesssim (q+q'-1)^2\mathcal I
\end{equation}
with probability larger than $1-O((L+n)^{-\nu})$ for some constant $\nu>0$. Moreover, with the same probability,
\begin{equation}
\phantomsection
\label{nondebias1}
\|\tilde{M}-Q\|_2\lesssim (q+q'-1)^2\mathcal I+\|\tilde{Q}-Q\|_2.
\end{equation}
\end{proposition}

The proof is similar to that of Theorem \ref{centralbound} provided in Appendix \ref{app::proof}; hence we omit it. It can be easily seen that $\mathcal I$ is larger than (\ref{bound1}) in Theorem \ref{centralbound}. Hence, if $q,q'>c$ for some $c>1/2$, which implies that the multiplier $(q+q'-1)^2$ is a constant, then \eqref{nondebias} and \eqref{nondebias1} are both inferior than (\ref{bound1}). To derive the bound for the eigen-vector and the misclassification error bound, we note that we can either use (\ref{nondebias}) or (\ref{nondebias1}). What we need is only the minimum non-zero eigen-value of the corresponding population matrix. Specifically, if we use (\ref{nondebias1}), then the population matrix is $Q$, which is the same as our setting. While if we use (\ref{nondebias}), then the minimum non-zero eigen-value of $\tilde{Q}$ is required. Here, we consider the two-parameter SBMs that $P_l$'s are homogeneous, with the inter-community and cross-community probabilities being $a$ and $b$, respectively. In this regime, the minimum non-zero eigen-value of $(q+q'-1) P_l+(1-q')\textbf{1}_n \textbf{1}_n^\T$ is identical to that of $(q+q'-1) P_l$, which implies that the error for the eigen-vector (counterpart of Theorem \ref{sparsecor}) and the misclassification error (counterpart of Theorem \ref{localmis}) totally depend on (\ref{nondebias1}). As a result, we can also see the accuracy gain in eigen-vector estimation and clustering.

\section{Additional numerical analysis and results}
\phantomsection
\label{app::exp}

\subsection{Effect of heterogeneity}
We conduct the following simulation to show the effect of heterogeneity.
\paragraph{Network-generation.} In Theorem \ref{oneshot-theory}, we show that the eigen-gap (i.e., $\lambda_{{\min}}(Q)$) divided by the heterogeneity of connectivity matrices (i.e., $\mathcal H$ in Definition \ref{balb}) can affect the estimation accuracy of the proposed algorithm theoretically. The heterogeneity can lead to either accuracy gain or loss, depending on the model structure of the network. To illustrate this phenomenon numerically, we consider the following two models for demonstrating the effect of heterogeneity.
\begin{itemize}
\item \emph{Model I.}
We fix $K=2$, $L=m=12$, $q,q'=0.8$ and let $n$ vary from 20 to 400. The communities are balanced. The networks are generated from multi-layer SBMs and equally belong to the following four different classes,
\begin{equation*}B^{(i)}=0.25\cdot{\diag}(1,1)+\alpha C^{(k)}
\end{equation*}
with
{\scriptsize
\begin{align*}
&C^{(1)}=
\begin{pmatrix}
0.1&0\\
0&0.1
\end{pmatrix},\quad
C^{(2)}=-
\begin{pmatrix}
0.1&0\\
0&0.1
\end{pmatrix},\quad
C^{(3)}=-
\begin{pmatrix}
0.1&0\\
0&0
\end{pmatrix},\quad
C^{(4)}=-
\begin{pmatrix}
0&0\\
0&0.1
\end{pmatrix}.
\end{align*}}\\
For this model, one can imagine that larger $\alpha$ indicates larger heterogeneity but a similar level of signal strength (eigen-gap). Therefore, larger $\alpha$ leads to larger effective heterogeneity (i.e., eigen-gap divided by heterogeneity; see the discussions after Theorem \ref{oneshot-theory} for the details). This is verified in Figure \ref{curseofhet}(a).

\item \emph{Model II.} We fix $K=2$, $L=m=12$, $q,q'=0.8$ and let $n$ vary from 20 to 400. The communities are balanced. Similar to \citet{arroyo2021inference}, the networks are generated from SBMs and equally belong to the following four different classes, \begin{equation*}
B^{(i)}=0.25( 11^\T)+\alpha C^{(k)}
\end{equation*}
with
{\scriptsize
\begin{align*}
&C^{(1)}=
\begin{pmatrix}
0.1&0\\
0&0.1
\end{pmatrix},
\quad
C^{(2)}=-
\begin{pmatrix}
0.1&0\\
0&0.1
\end{pmatrix},\quad
C^{(3)}=
\begin{pmatrix}
0.1&0\\
0&0
\end{pmatrix},
\quad
C^{(4)}=
\begin{pmatrix}
0&0\\
0&0.1
\end{pmatrix}.
\end{align*}
}\\
For this model, larger $\alpha$ indicates larger heterogeneity but stronger signal strength (eigen-gap). Thus, the composite effects lead to larger $\alpha$ corresponding to smaller effective heterogeneity. See Figure \ref{blessofhet}(a) for the illustration.
\end{itemize}

\paragraph{Results.} The averaged results over 20 replications for \emph{Model I} and \emph{II} are displayed in Figure \ref{curseofhet} and \ref{blessofhet}, respectively. The results are consistent with the theoretical results given in Theorem \ref{oneshot-theory}.
For \emph{Model I}, we see that large heterogeneity (i.e., large $\alpha$) leads to inferior performance of \texttt{ppDSC}, especially in terms of the projection distance, which is partially because large $\alpha$ leads to large effective heterogeneity (i.e., eigen-gap divided by heterogeneity) in Theorem \ref{oneshot-theory}; see Figure \ref{curseofhet}. For \emph{Model II}, the results are on the opposite side in that large heterogeneity (i.e., large $\alpha$) leads to performance gain of \texttt{ppDSC}, which is partially due to the fact that large $\alpha$ leads to large eigen-gap and small effective heterogeneity; see Figure \ref{blessofhet}. In addition, our estimator \texttt{ppDSC} is better than the biased \texttt{ppDSC-1b} and \texttt{ppDSC-2b}, and is closer to the baseline estimators \texttt{ppSC} and \texttt{Oracle} estimator.
\begin{figure*}[h]{}
\centering
\subfigure[Population parameters]{\includegraphics[height=4.5cm,width=4.5cm,angle=0]{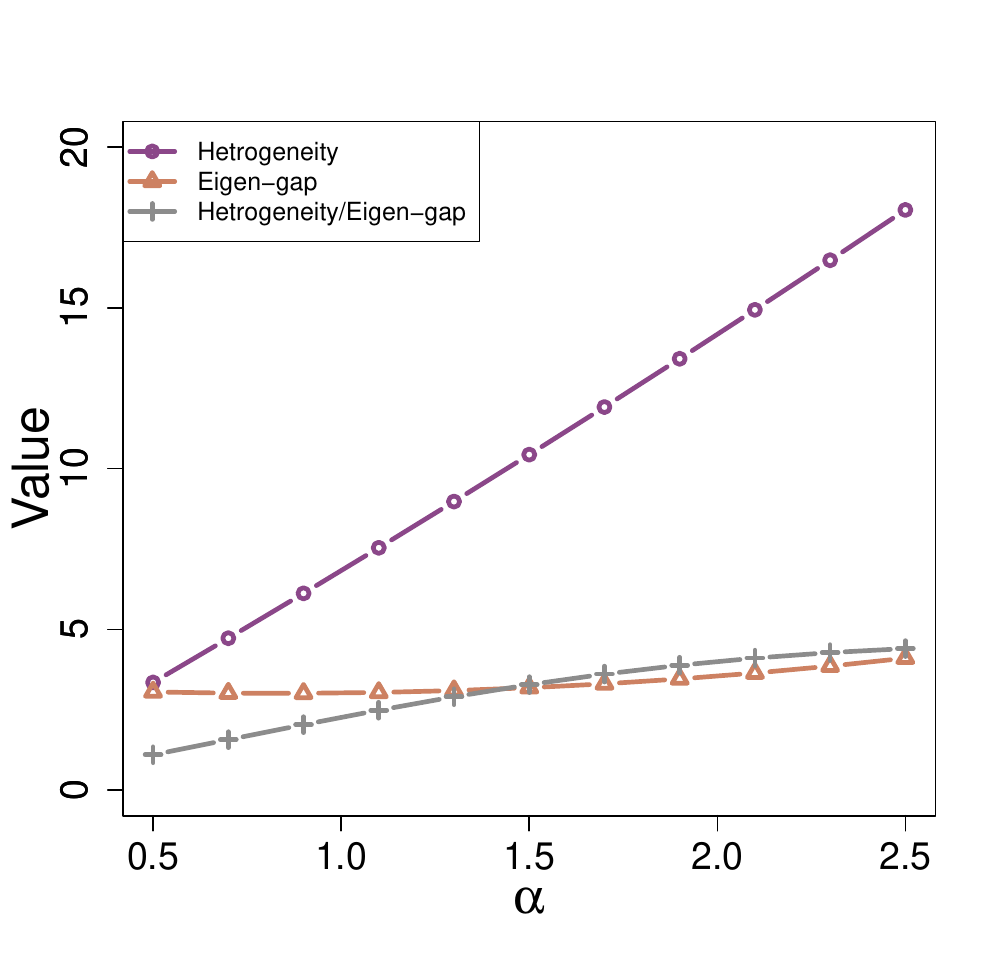}}
\subfigure[Projection distance]{\includegraphics[height=4.5cm,width=4.5cm,angle=0]{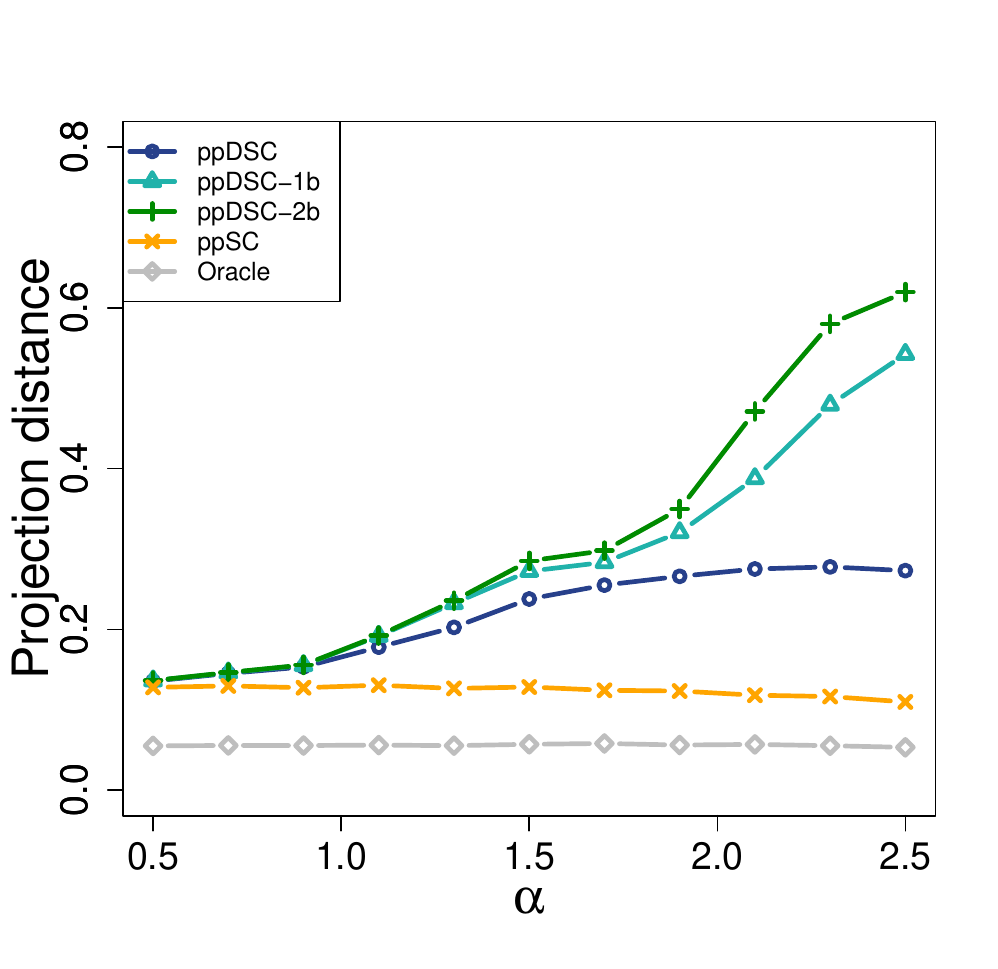}}
\subfigure[Misclassification rate]{\includegraphics[height=4.5cm,width=4.5cm,angle=0]{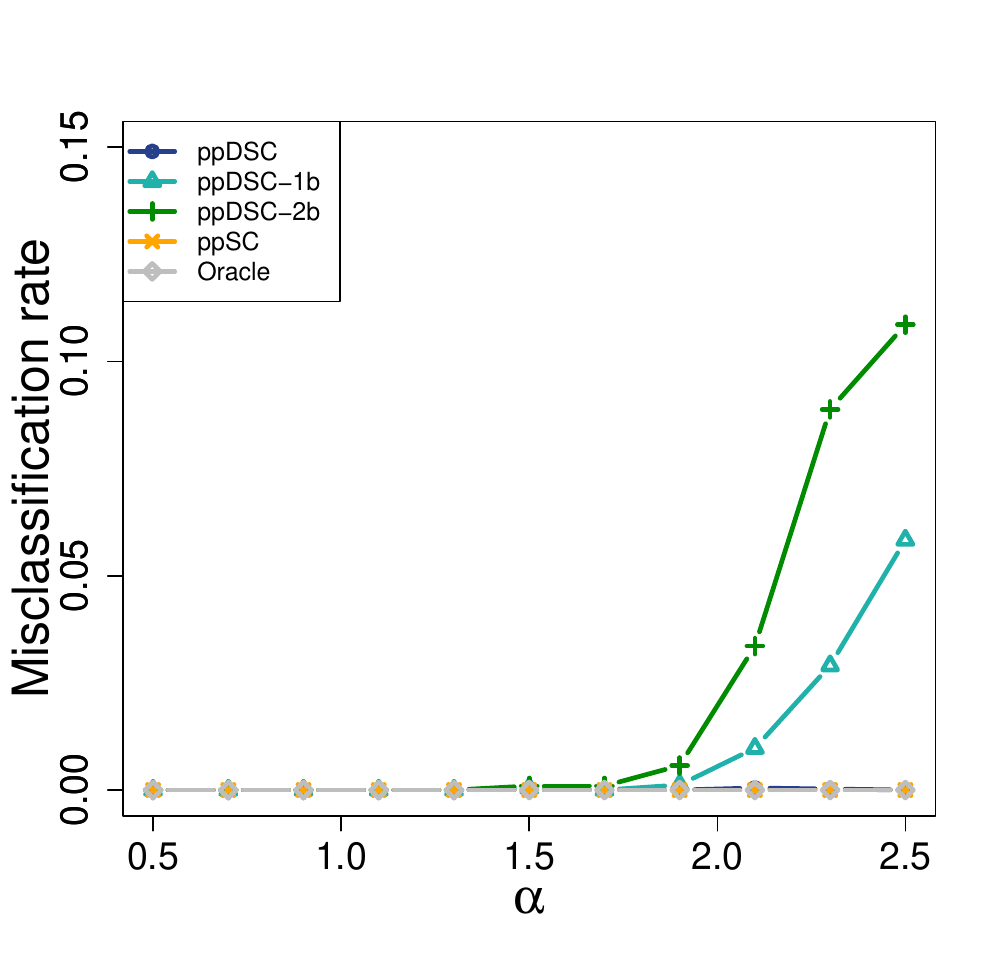}}
\caption{Illustration for the negative effect of heterogeneity (\emph{Model I}). (a) corresponds to population parameters including the heterogeneity, eigen-gap and their ratios against $\alpha$. (b) and (c) correspond to the projection distance and misclassification rate of \texttt{ppDSC}, \texttt{ppDSC-1b}, \texttt{ppDSC-2b}, \texttt{ppSC} and \texttt{Oracle}. }\label{curseofhet}
\end{figure*}

\begin{figure*}[!htbp]{}
\centering
\subfigure[Population parameters]{\includegraphics[height=4.5cm,width=4.5cm,angle=0]{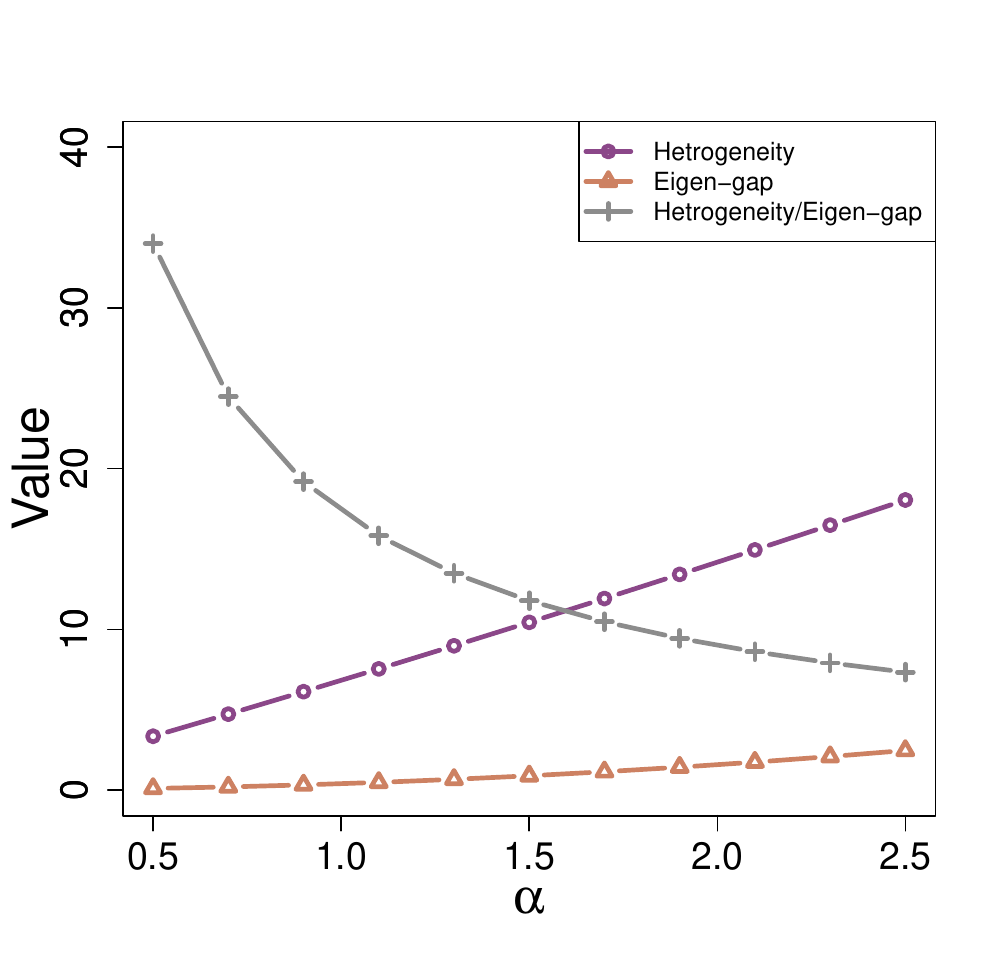}}
\subfigure[Projection distance]{\includegraphics[height=4.5cm,width=4.5cm,angle=0]{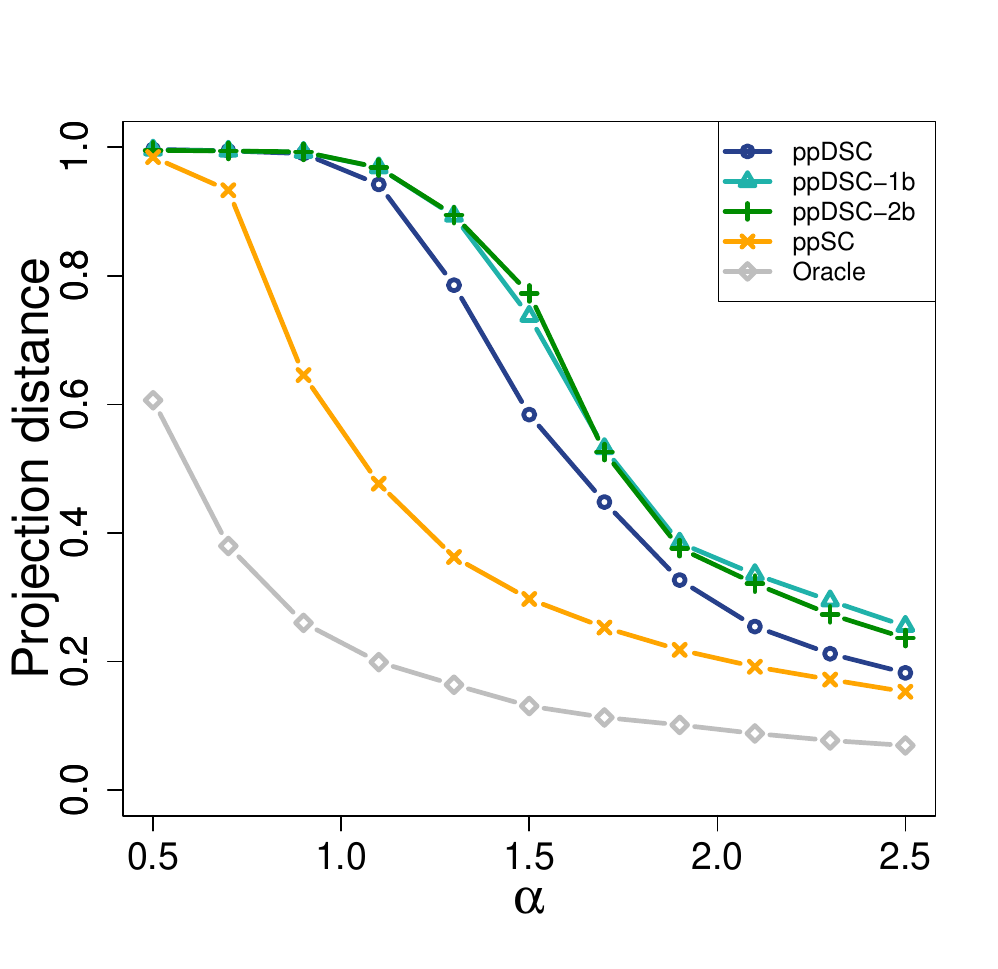}}
\subfigure[Misclassification rate]{\includegraphics[height=4.5cm,width=4.5cm,angle=0]{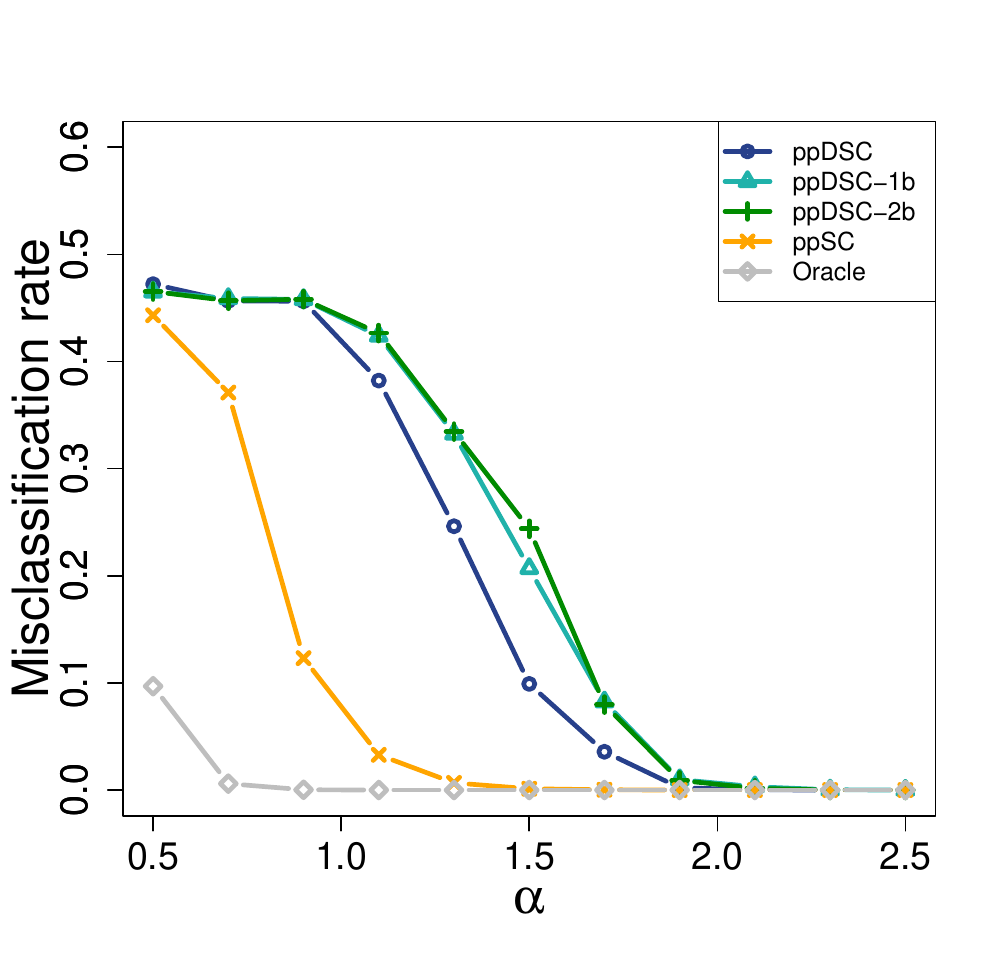}}
\caption{Illustration for the positive effect of heterogeneity (\emph{Model II}). (a) corresponds to population parameters including the heterogeneity, eigen-gap and their ratios against $\alpha$. (b) and (c) correspond to the projection distance and misclassification rate of \texttt{ppDSC}, \texttt{ppDSC-1b}, \texttt{ppDSC-2b}, \texttt{ppSC} and \texttt{Oracle}.}\label{blessofhet}
\end{figure*}

\subsection{Effect of network sparsity}
\textcolor{black}{We evaluate the effect of network sparsity on the performance of \texttt{ppDSC} under different privacy budgets. To incorporate the network sparsity, we generate $B_l$'s by
$B_l:=0.8\gamma B^{(1)}$ for $l=1,...,\lfloor L/2\rfloor$ and $B_l:=0.6\gamma B^{(2)}$ for $l=\lfloor L/2\rfloor+1,...,L$ with varying $\gamma$.
Other parameters are $n=210$ and $L=m=12$. The averaged results over 20 replications under different $q$'s are displayed in Figure \ref{sparsityvaluation}. We assume $q,q'$ relates to $\epsilon$ via $q=q'=\frac{{\rm e}^\epsilon}{1+{\rm e}^\epsilon}$. Under a given privacy budget, a denser network yields better clustering performance. In other words, under a given network sparsity, a denser network permits a wider range of privacy budget to attain a specified misclassification rate or projection distance.}

\begin{figure*}[!h]{}
\centering
\subfigure[Projection distance]{\includegraphics[height=5.5cm,width=5.6cm,angle=0]{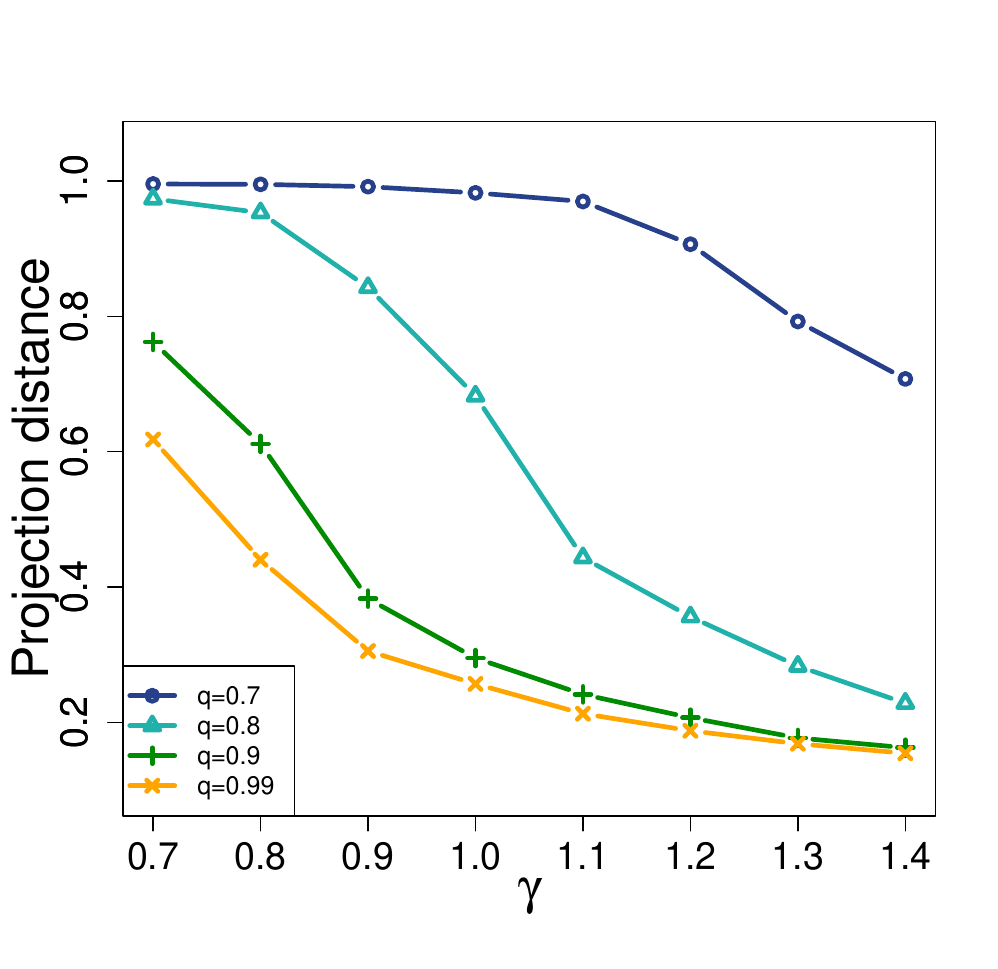}}\hspace{0.3cm}
\subfigure[Misclassification rate]{\includegraphics[height=5.5cm,width=5.6cm,angle=0]{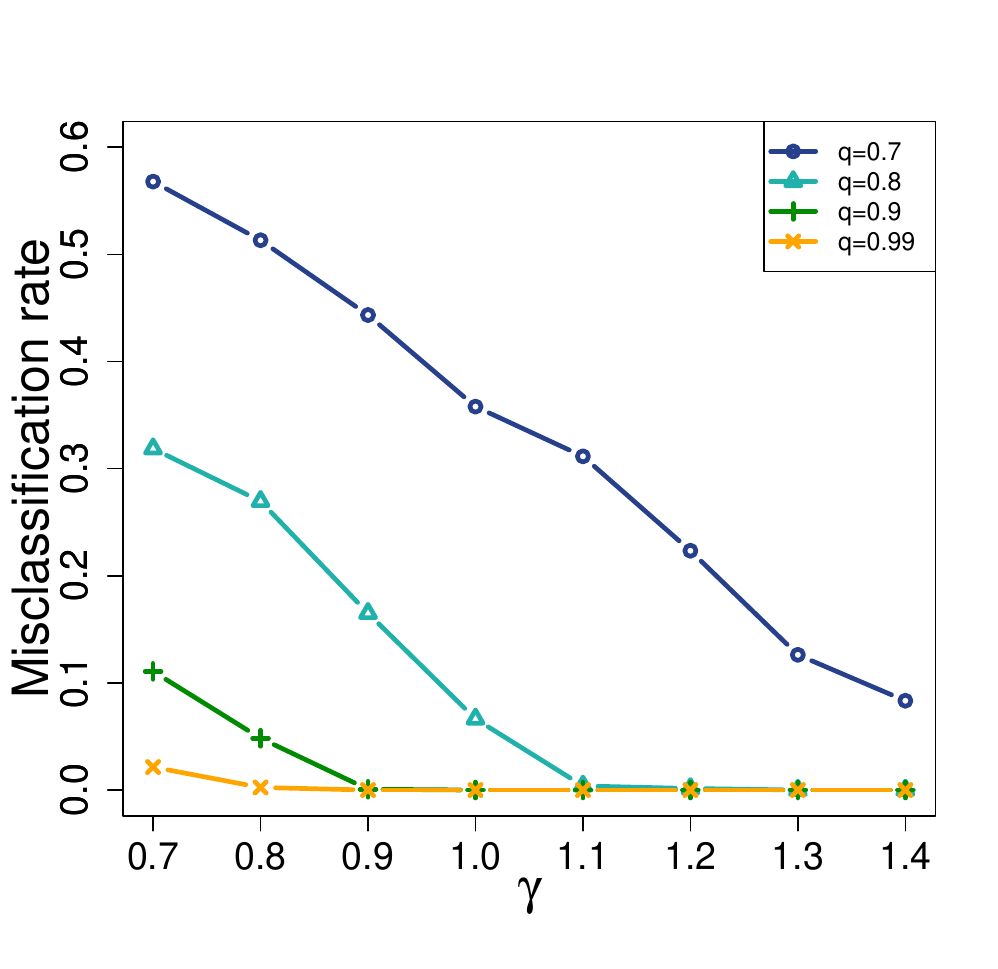}}
\caption{The performance of \texttt{ppDSC} with varying sparsity under different $q,q'$'s.}\label{sparsityvaluation}
\end{figure*}

\subsection{Experimental details and additional figures of real data analysis}
Figures \ref{localnetwork} and \ref{localmatrix} show the visualization of the \texttt{AUCS} dataset in terms of networks and matrices, respectively.

\begin{figure*}[!h]{}
\centering
\subfigure[Work]{\includegraphics[height=4.4cm,width=4.4cm,angle=0]{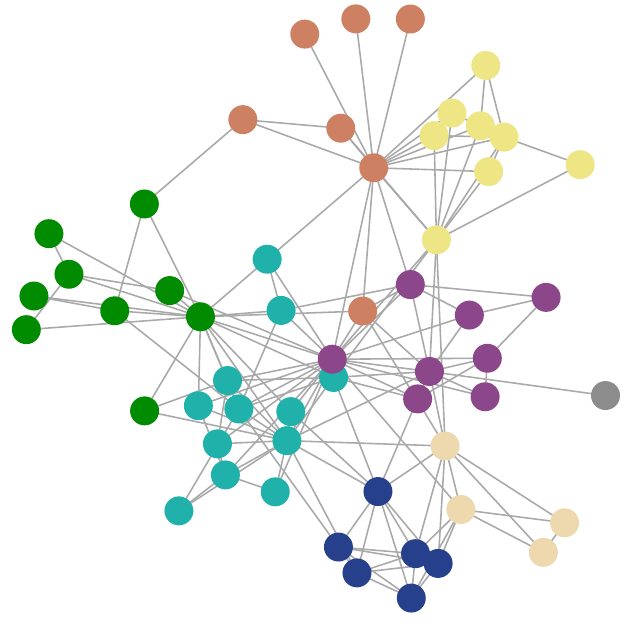}}
\subfigure[Facebook]{\includegraphics[height=4.4cm,width=4.4cm,angle=0]{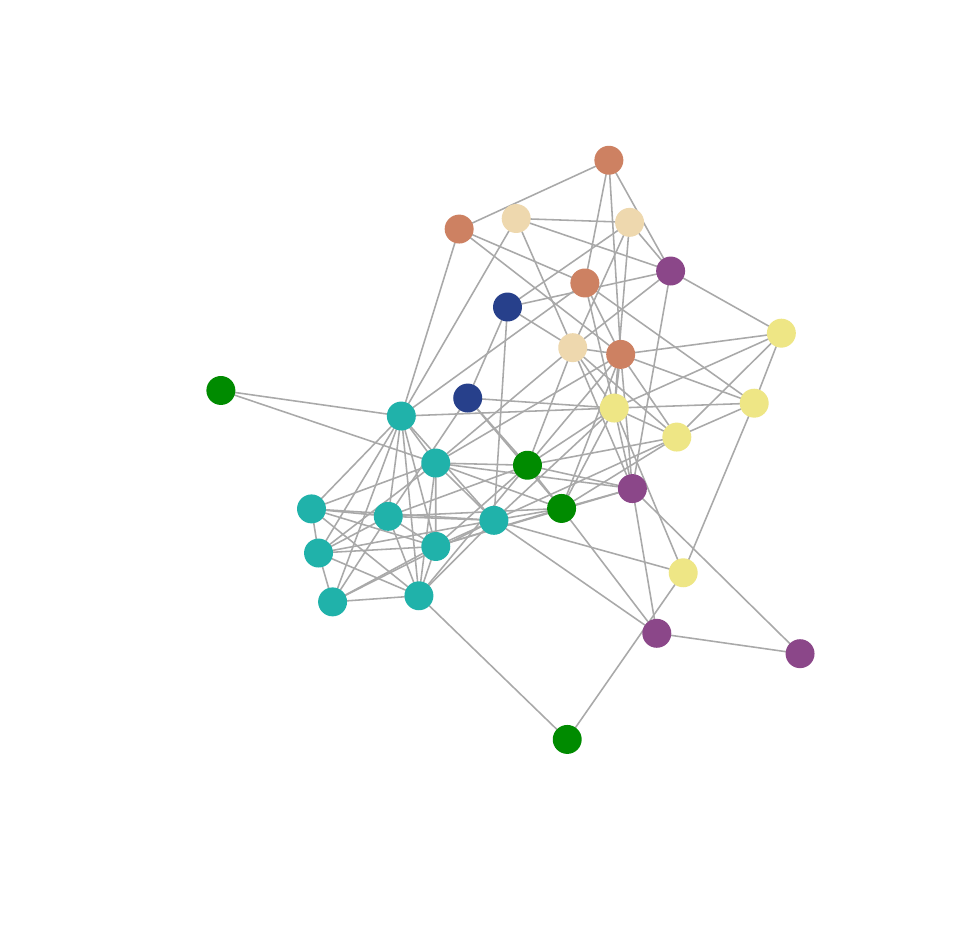}}
\subfigure[Leisure]{\includegraphics[height=4.4cm,width=4.4cm,angle=0]{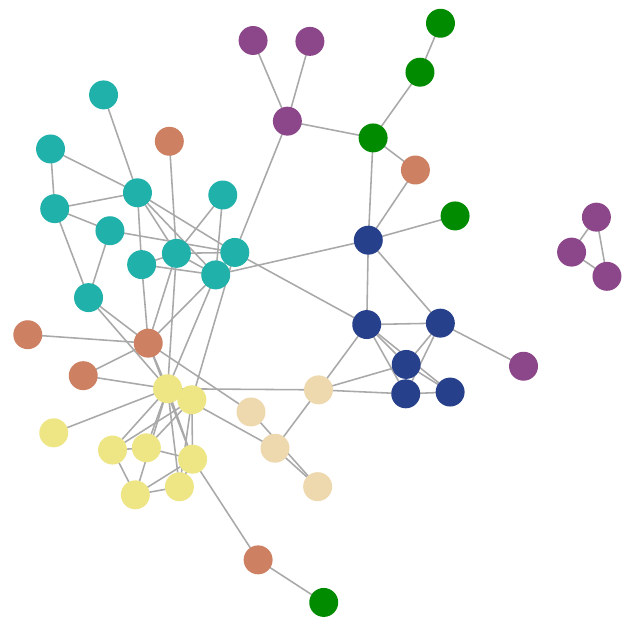}}
\subfigure[Lunch]{\includegraphics[height=4.4cm,width=4.4cm,angle=0]{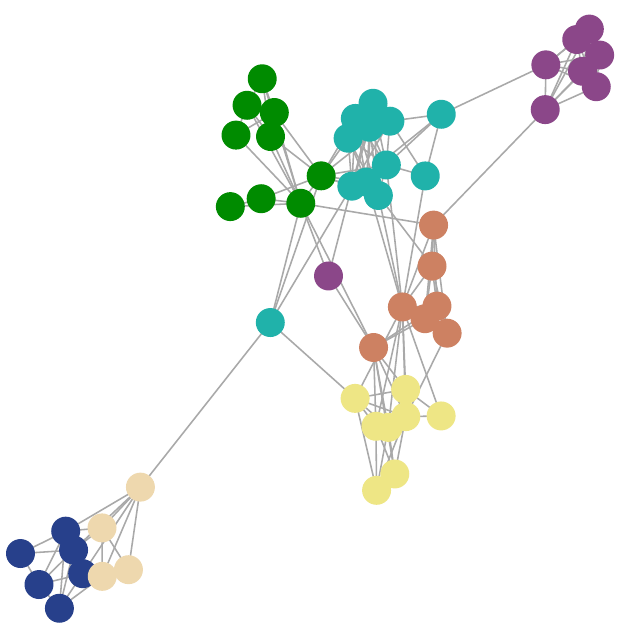}}
\subfigure[Coauthor]{\includegraphics[height=4.4cm,width=4.4cm,angle=0]{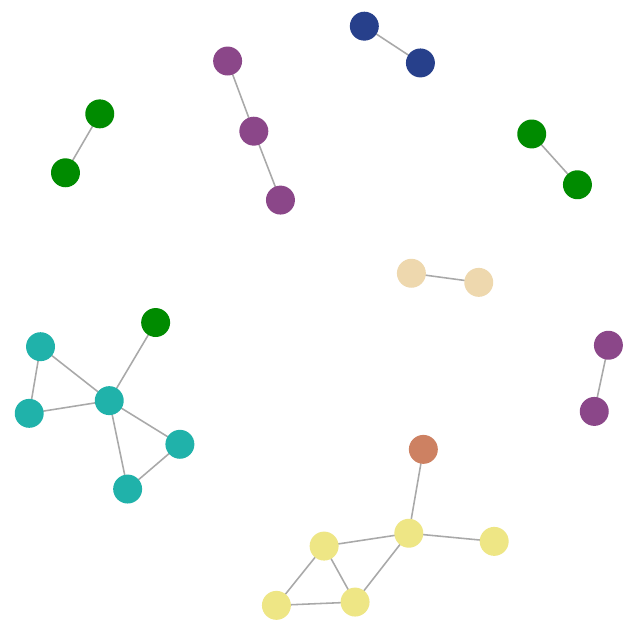}}
\caption{A visualization of the \texttt{AUCS} network where vertex colors represent research groups. Each network corresponds to a specific relationship among the vertices.}\label{localnetwork}
\end{figure*}

\begin{figure*}[!h]{}
\centering
\subfigure[Whole]{\includegraphics[height=4.2cm,width=4.4cm,angle=0]{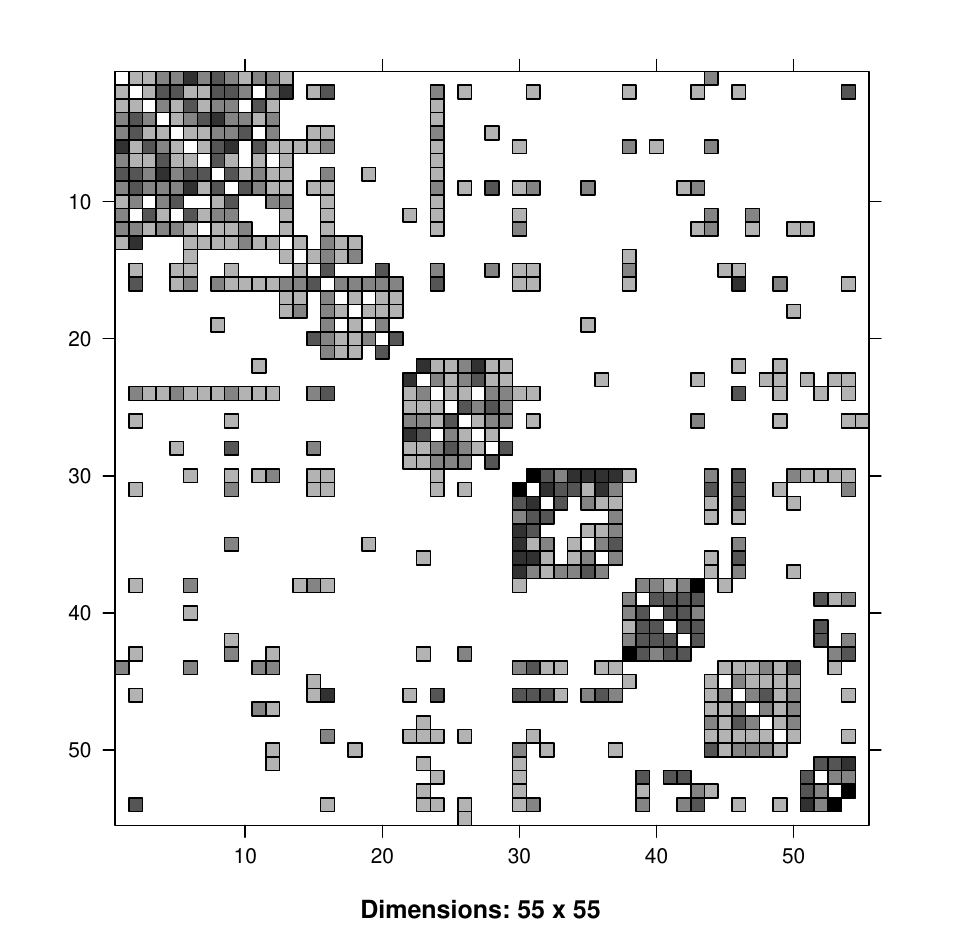}}
\subfigure[Work]{\includegraphics[height=4.2cm,width=4.4cm,angle=0]{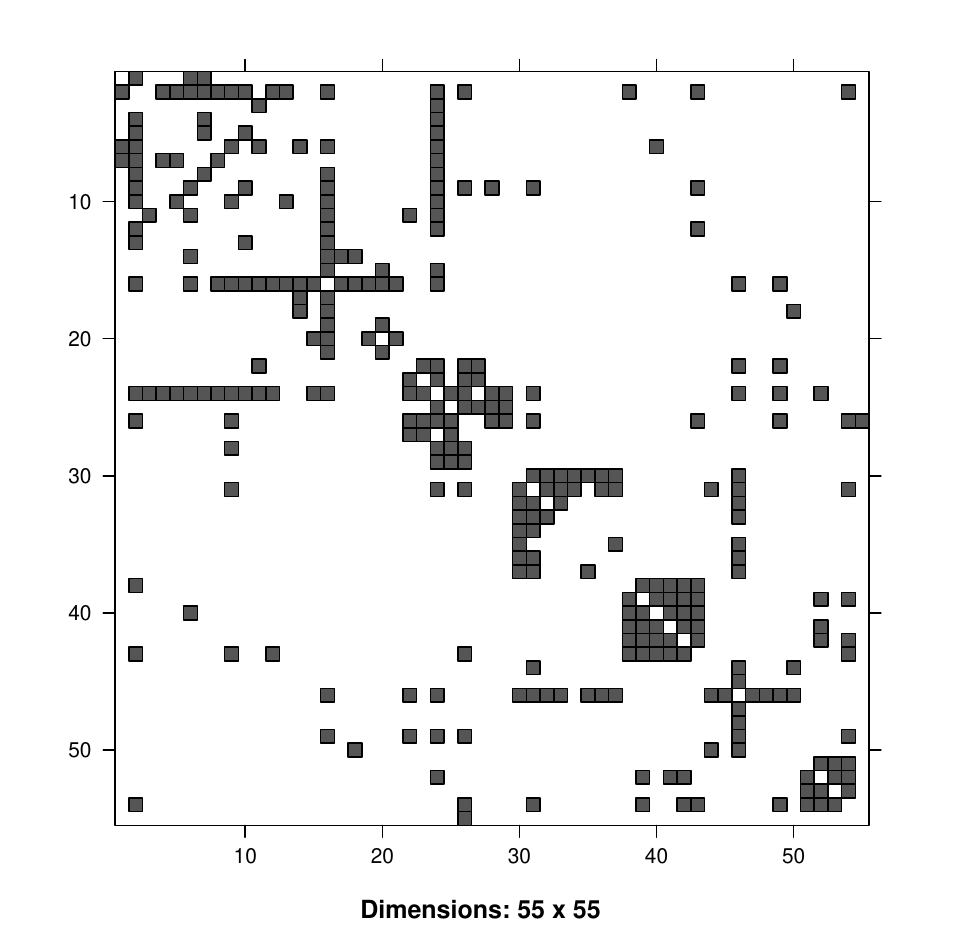}}
\subfigure[Facebook]{\includegraphics[height=4.2cm,width=4.4cm,angle=0]{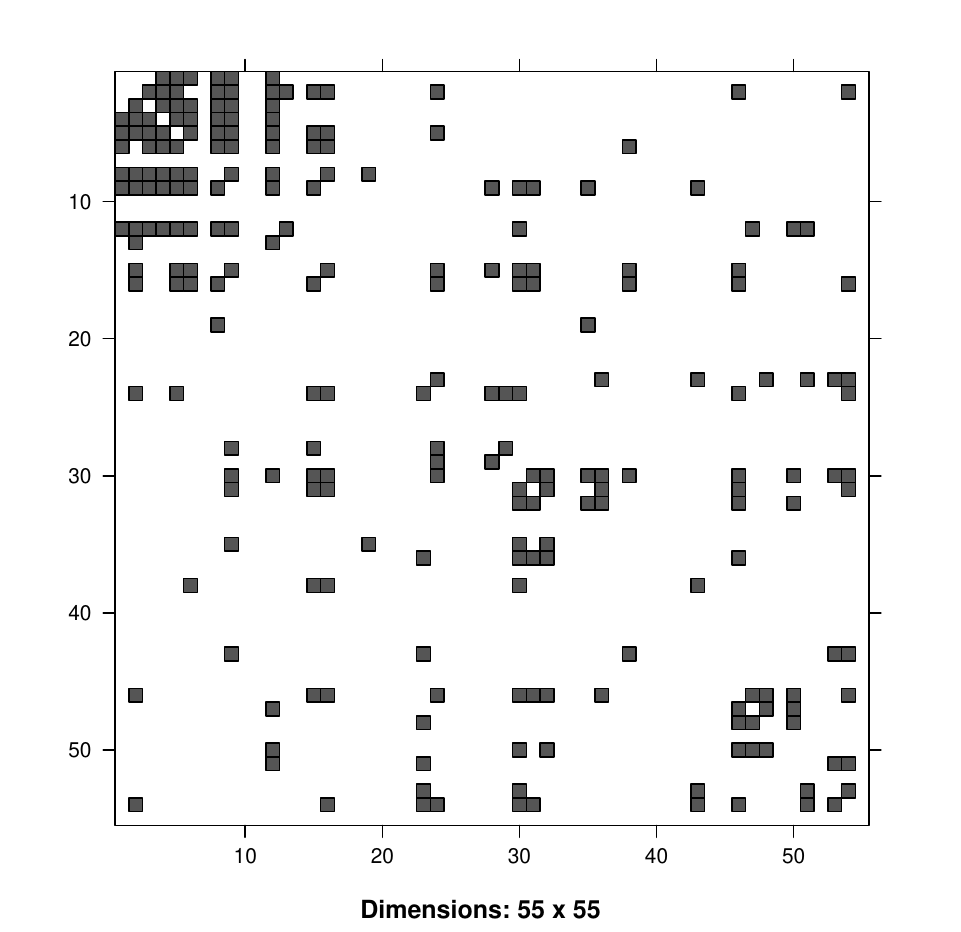}}
\subfigure[Leisure]{\includegraphics[height=4.2cm,width=4.4cm,angle=0]{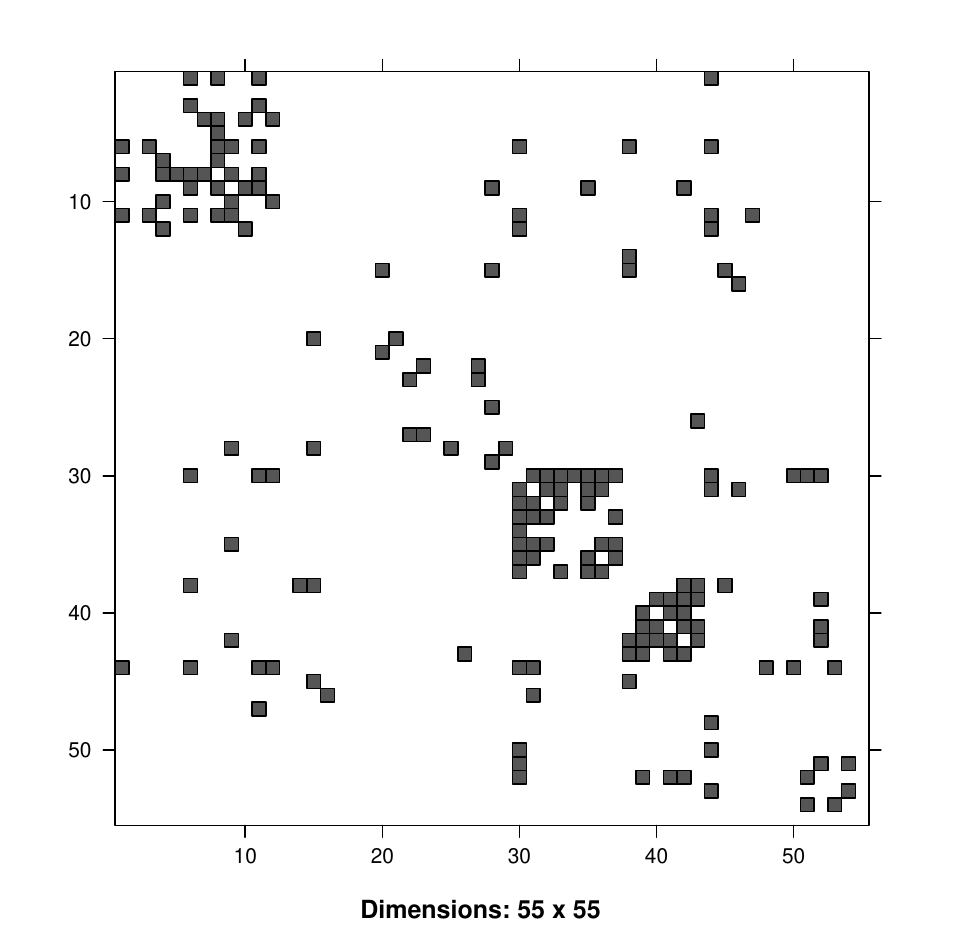}}
\subfigure[Lunch]{\includegraphics[height=4.2cm,width=4.4cm,angle=0]{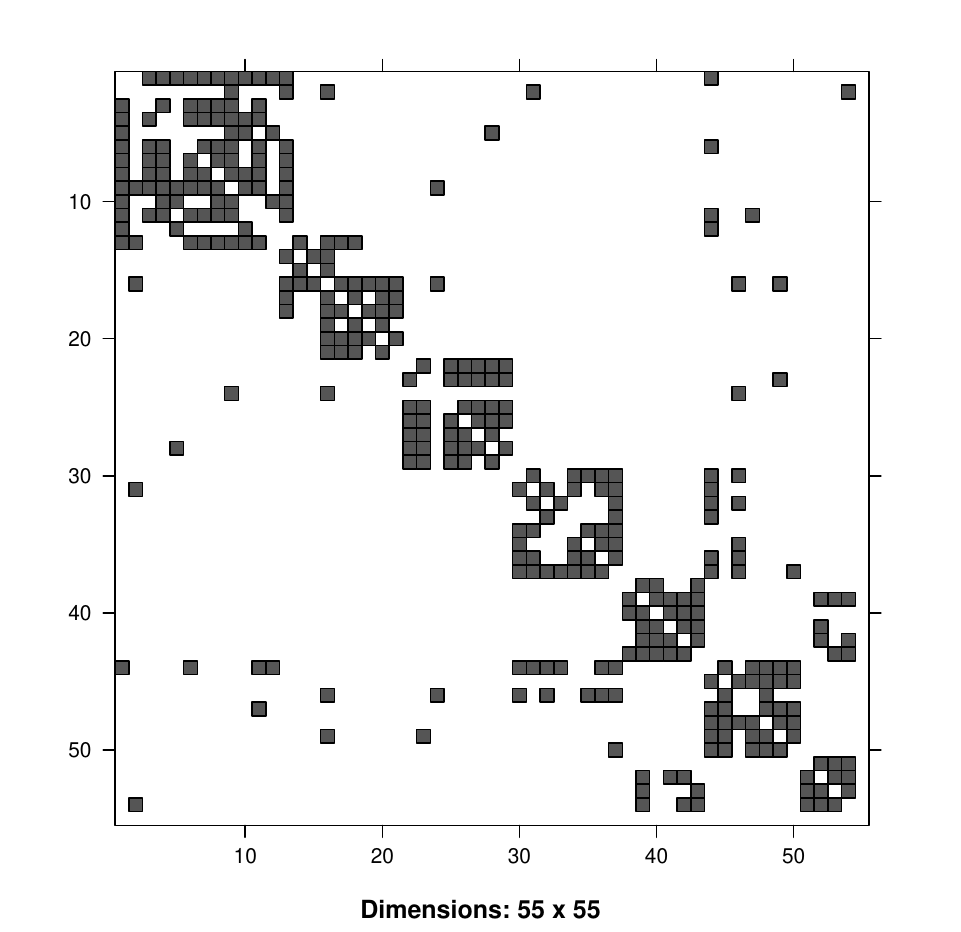}}
\subfigure[Coauthor]{\includegraphics[height=4.2cm,width=4.4cm,angle=0]{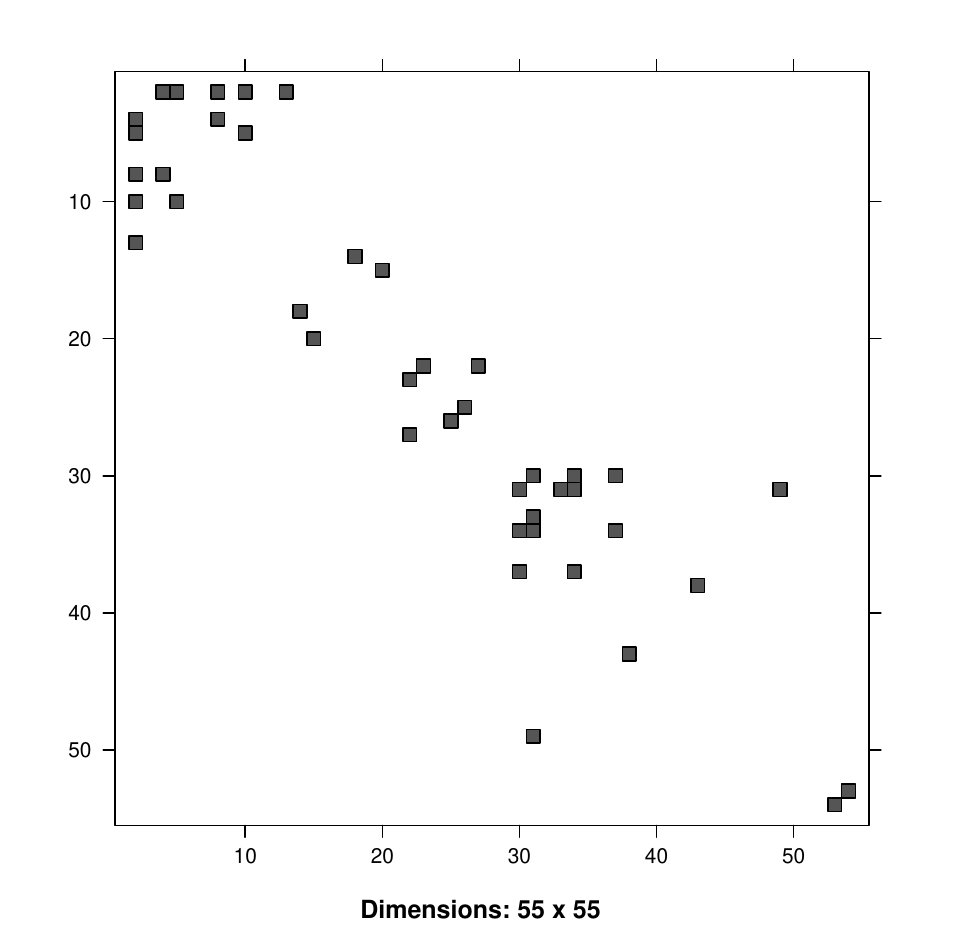}}
\caption{A visualization of the network adjacency matrices of the \texttt{AUCS} dataset. The nodes are permutated according to their underlying research groups. (a) corresponds to the network adjacency matrix that contains all the edges of the five relationships. (b)-(f) correspond to the adjacency matrices of the five relationships among vertices, respectively. }\label{localmatrix}
\end{figure*}

We next provide the details of the methods compared in Section \ref{sec::real}.

\paragraph{Methods compared in Section \ref{sec::real}.} Consistent with the simulation experiments, we compare our method \texttt{ppDSC} with \texttt{ppDSC-1b}, \texttt{ppDSC-2b} and \texttt{Oracle}. The detailed description of each method can be found in Section \ref{sec:sim}. Generally, we study how the performance of each method alters with the network number $L$ and the privacy parameters $q$ and $q'$. Regarding the effect of $L$, we mean that we use some of the five local networks to construct the estimator to see how the number of used networks affects the clustering performance. Regarding the effect of $q,q'$, we consider three cases, including $q,q'$ both varying, $q$ varying but $q'$ being fixed, and $q'$ varying being $q$ being fixed, where the latter two cases aim to verify the different effect of $q$ and $q'$ indicated by the theory.
The detailed setups are as follows.
\begin{itemize}
\item  Effect of $L$: $q=q'=0.9$, and $L$ ranges from 2 to 5.
\item  Effect of $q,q'$: $L=5$, and $q$ and $q'$ are equal and range from 0.6 to 1.
\item  Effect of $q$: $L=5$, $q'=0.95$, and $q$ ranges from 0.6 to 1.
\item  Effect of $q'$: $L=5$, $q=0.95$, and $q'$ ranges from 0.6 to 1.
\end{itemize}

\section{Main proofs}
\phantomsection
\label{app::proof}
\subsection{Proof of Theorem \ref{privacy-theory} }
For $l\in \{1,...,L\}$, denote the RR-perturbed $A_i$ by $\tA_i$. By Proposition \ref{dp-prop}, for each $i$, $\tA_i$ satisfies $\epsilon$-edge-DP. Further, by the parallel composition theorem \citep{mcsherry2009privacy}, $(\tA_1,...,\tA_L)$ also satisfies $\epsilon$-edge-DP. Finally, by the post-processing property of DP (Proposition \ref{dp-prop}), Algorithm \ref{Oneshotsvd} can achieve $\epsilon$-edge-DP.\QEDA

\subsection{Proof of Theorem \ref{oneshot-theory}}
First, we note that we have a degree of freedom in choosing $V$. For simplicity, we assume that the reference solution $\hat{V}_o$ in Algorithm \ref{Oneshotsvd} is already aligned with $V$ in that $\arg \min_{Z\in\mathcal O_K} \|\hat{V}_oZ-V\|_F=I_K$. Otherwise, we could work with $VZ$ for some $Z\in \mathcal O_K$ instead.

We will make use of the following result to prove, which is modified from \citet{charisopoulos2021communication} using the notation of this work.

\begin{proposition}
\label{oneshotoriginal}
  Suppose
  \begin{equation}
  \label{con1}
  \lambda_{{\min}} (Q)\geq \delta,\quad
  {for \; some \;} \delta>0
  \end{equation}
 and
  \begin{equation}
  \label{con2}
  \|\sum_{l\in S_i}\overline{M}_l/(L/m)-Q\|_2\leq \frac{\delta}{8},\quad{for \; all \;} i=1,...,m.
  \end{equation}
 Then the output $\tilde{V}$ of Algorithm \ref{Oneshotsvd} satisfies that
  \begin{equation*}
  {\mathrm{dist}}(\tilde{V},V)\lesssim \frac{1}{\delta^2}\max_{i\in [m]} \|\sum_{l\in S_i}{\overline{M}_l}/{(L/m)}-Q\|_2^2+\frac{1}{\delta}\|\frac{1}{L}\sum_{l=1}^L \overline{M}_l-Q\|_2.
  \end{equation*}
\end{proposition}

We first have that
\begin{align*}
\max_{i\in [m]} \|\sum_{l\in S_i}{\overline{M}_l}/{L/m}-Q\|_2&\leq \max_{i\in [m]} \|\sum_{l\in S_i}{(\overline{M}_l-Q_l)}/{(L/m)}\|_2+\max_{i\in [m]} \|\sum_{l\in S_i}{Q_l}/{(L/m)}-Q\|_2\nonumber\\
&\lesssim  \max_{i\in [m]} \|\sum_{l\in S_i}{(\overline{M}_l-Q_l)}/{(L/m)}\|_2+\mathcal H(n,\rho, L),
\end{align*}
where the last inequality follows from
\begin{align*}
\phantomsection
		\max_{i\in [m]} \|\sum_{l\in S_i}{Q_l}/{(L/m)}-Q\|_2&=\max_{i\in [m]}\left\|\frac{1}{L/m}\sum_{l\in S_i}\frac{P_l^2}{n}-\frac{1}{L}\sum_{l=1}^L \frac{P_l^2}{n}\right\|_2\\
  &\leq \max_{i\in [m]}\frac{1}{n}\|\Delta\|_2^2 \left\|\frac{1}{L/m}\sum_{l\in S_i}B_l\Delta ^2B_l-\frac{1}{L}\sum_{l\in [L]}B_l\Delta^2 B_l\right\|_2\nonumber\\
&\lesssim\max_{i\in [m]}\frac{1}{K} \left\|\frac{1}{L/m}\sum_{l\in S_i}B_l\Delta ^2B_l-\frac{1}{L}\sum_{l\in [L]}B_l\Delta^2 B_l\right\|_2\asymp \mathcal H(n,\rho, L),
\end{align*}
by using the fact that
$$P_l^2=\Theta B_l \Theta^\T \Theta B_l \Theta^\T =\bar{\Theta} \Delta (B_l \Delta^2 B_l) \Delta\bar{\Theta}^\T$$
with $\Delta:={\diag}(\sqrt{n_1},...,\sqrt{n_{K}})$, $\bar{\Theta}:=\Theta \Delta^{-1}$, and $\bar{\Theta}^\T \bar{\Theta}=\mathbb I$,
and Assumption \ref{balcom} and Definition \ref{balb}.
To meet (\ref{con1}) and (\ref{con2}), we let $\delta\asymp \lambda_{\min} (Q)$ and enforce
$$ \max_{i\in [m]} \|\sum_{l\in S_i}(\overline{M}_l-Q_l)/{(L/m)}\|_2\lesssim\frac{\lambda_{\min} (Q)}{16}\quad {and} \quad\mathcal H\lesssim \frac{\lambda_{\min} (Q)}{16},$$
which follows by our condition in Theorem \ref{oneshot-theory} that $\mathcal I_1$ and $\mathcal I_2$ are smaller than a small constant $c$. We can thus use the results in Proposition \ref{oneshotoriginal} to obtain that,
\begin{equation*}
\mathrm{dist} (\tilde{V}, V)\lesssim {\max_l\frac{\|\sum_{l\in S_i}(\overline{M}_l-Q_l)/{(L/m)}\|_2^2}{\lambda_{\min}^2 (Q)}}+{\frac{\mathcal H^2(n,\rho, L)}{\lambda_{\min}^2 (Q)}}+\frac{\|\sum_l\overline{M}_l/L-Q\|_2}{\lambda_{\min}(Q)}.
\end{equation*}
The results of Theorem \ref{oneshot-theory} are proved.
\QEDA

\subsection{Proof of Theorem \ref{centralbound}}
We make use of the concentration tools developed in \citet{lei2020bias} to prove.
Recall that $\overline{M}=\frac{1}{L}\sum_{l=1}^L \overline{M}_l$ and (\ref{secdb}), then we can decompose the difference between $\overline{M}$ and $Q$ as
\begin{equation*}
\phantomsection
n\overline{M}-nQ=\sum_{l=1}^L \frac{1}{L}(n\overline{M}_l-nQ_l):=E_0+E_1+E_2+E_3-\sum_{l}\frac{1}{L}\frac{G_l(q')^2}{(q+q'-1)^2},
\end{equation*}
with $E_{i}:= \frac{1}{L}\sum_{l=1}^L E_{l,i}$ for $i=0,1,2,3$ and $E_{l,i}$ is the same with those in (\ref{composion2}). Now, we bound these error terms, respectively.

For $E_{0}:=\sum_{l}\frac{1}{L}\left({-P_l}\cdot{\diag}(P_l)-{\diag}(P_l)\cdot P_l +({\diag}(P_l))^2\right)$, we have
\begin{align*}
\|E_{0}\|_2\leq 2\max_l\|{\diag}(P_l)\|_2\max_l \|P_l\|_{{F}}+\max _l\|{\diag}(P_l)\|_2^2\leq Cn\rho^2.
\end{align*}

For $E_{1}:=\sum_{l}\frac{1}{L}\left(X_l (P_l-\diag(P_l))+ (P_l-\diag(P_l))X_l\right)$, we would use the result in Lemma \ref{lei-bound1}. To that end, we first show that each entry of $(q+q'-1)X_l$ satisfies the Bernstein tail condition (see Definition \ref{bernstein}). Given $i\neq j$, with slight abuse of notation, define $$Y_l:=(q+q'-1)X_{l,ij}=(q+q'-1)(\overline{A}_{l,ij}-P_{l,ij})=\tA_{l,ij}-(1-q')-(q+q'-1)P_{l,ij}.$$ Then
\begin{align}
Y_l=\begin{cases}
q'-(q+q'-1)P_{l,ij}\;\, \mathrm{with\; probability}\, \; qP_{l,ij}+(1-P_{l,ij})(1-q'),\nonumber\\
q'-1-(q+q'-1)P_{l,ij}\;\, \mathrm{with\; probability} \,\; q'(1-P_{l,ij})+P_{l,ij}(1-q),\nonumber
\end{cases}
\end{align}
and for $k\geq 2$, we have
\begin{align}
\label{bern}
\ME |Y_l^k|&\leq |q'-(q+q'-1)P_{l,ij}|^k\Big[qP_{l,ij}+(1-P_{l,ij})(1-q')\Big]\nonumber\\
&\quad\quad\quad+|(q+q'-1)P_{l,ij}+1-q'|^k \Big[q'(1-P_{l,ij})+P_{l,ij}(1-q)\Big]\nonumber\\
&\leq \left(qP_{l,ij}+(1-P_{l,ij})(1-q')\right)+\left((q+q'-1)P_{l,ij}+1-q'\right)\nonumber\\
&\leq 2\left(q\rho+1-q'\right)\leq \frac{v_1}{2}k!R_1^{k-2},
\end{align}
where $v_1:=4\left(q\rho+1-q'\right)$ and $R_1:=1$, and we used the fact that $P_{l,ij}=o(1)$. Hence, $(q+q'-1)X_{l,ij}$ is $(v_1,R_1)$-Bernstein. In addition, it holds that $\|\sum_l(P_l-{\diag}(P_l))^2\|_2\leq \sum_l\|P_l-{\diag}(P_l)\|_{{F} }^2\leq L n^2\rho^2$ and $\max_l\|P_l-{\diag}(P_l)\|_{2,\infty}\leq \sqrt{n}\rho$. Applying Lemma \ref{lei-bound1}, we obtain
\begin{align*}
\MP((q+q'-1)L\|E_{1}\|_2\geq t)\leq 4n\cdot \exp\left\{\frac {- t^2/2}{4(q\rho +(1-q'))Ln^3\rho^2+\sqrt{n}\rho t}\right\}.
\end{align*}
Choosing $t=c(q\rho+1-q')^{1/2}n^{3/2}L^{1/2}\rho \log^{1/2} (L+n)$, it turns out that
$$\sqrt{n}\rho t \leq c4(q\rho +(1-q'))L n^{3}\rho^2, $$
which is $\log^{1/2}(L+n)\leq c (q\rho+1-q')^{1/2}n L^{1/2}$ and can be certainly met if $\log^{1/2}(L+n)\leq c (1-q')^{1/2}n L^{1/2}$, which can be satisfied by the fact that $q'={\mathrm{e}}^\epsilon/(1+{\mathrm{e}}^\epsilon)$ and the condition (\ref{eandl}). As a result, we have with probability larger than $1-(L+n)^{-\beta}$ for some $\beta>0$ that
\begin{align*}
\|E_{1}\|_2\leq c \frac{(q\rho +1-q')^{1/2}}{q+q'-1}\cdot n^{3/2}\rho \log^{1/2} (L+n)/\sqrt{L}.
\end{align*}

For $E_{2}:=\sum_{l}\frac{1}{L}\left(X_l^2-{\diag}(X^2_l)\right)$, we would use the results in Lemma \ref{lei-bound2}. For this purpose, we first show that given an independent copy $(q+q'-1)\tilde{X}_l$ of $(q+q'-1)X_l$, each $(q+q'-1)^2X_{l,ij}\tilde{X}_{l,ij}$ satisfies the Bernstein tail condition. Actually, similar to the derivation of (\ref{bern}), we can easily show that $(q+q'-1)^2X_{l,ij}\tilde{X}_{l,ij}$ is $(v'_2,R'_2)$-Bernstein with $v'_2=8\left(q\rho+1-q'\right)^2$ and $R'_2=1$. Then by Lemma \ref{lei-bound2}, we obtain with probability larger than $1-O((L+n)^{-1})$ that
\begin{align}
\label{e2}
\|E_{2}\|_2 \leq c \max\left\{\frac{(q\rho+1-q')n\log(L+n),(q\rho+1-q')^{1/2}\sqrt{n}\log^{3/2} (L+n),\log^2(L+n)/\sqrt{L}}{\sqrt{L}(q+q'-1)^2}\right\}.
\end{align}
Further, by the condition (\ref{eandl}), we have {$\log(L+n)\leq c n(1-q')\leq c n(q\rho+1-q')$} and (\ref{e2}) reduces to
\begin{align*}
\|E_{2}\|_2 \leq c\frac{(q\rho+1-q')n\log (L+n)}{\sqrt{L}(q+q'-1)^2}.
\end{align*}

For $E_3-\sum_{l}\frac{1}{L}\frac{G_l(q')^2}{(q+q'-1)^2}$, {recalling (\ref{e3bound})}, it is easy to see that
\begin{align*}
\|E_3-\sum_{l}\frac{1}{L}\frac{G_l(q')^2}{(q+q'-1)^2}\|_2\leq C\frac{n((q+q'-1)\rho+1-q')^2}{(q+q'-1)^2}.
\end{align*}

Consequently, $E_1$, $E_2$ and $E_3$ are the main effects. Noting (\ref{simplenote}), we obtain the results of Theorem \ref{centralbound}.
\QEDA

\subsection{Proof of Theorem \ref{sparsecor}}
The proof of Theorem \ref{sparsecor} follows from that of Theorem \ref{oneshot-theory}. First, we need to show that $Q$ is of rank $K$ and provide the lower bound of $\lambda_{K}(Q)$. Indeed,
define $\Delta={\diag}(\sqrt{n_1},...,\sqrt{n_{K}})$ and denote $\bar{\Theta}:=\Theta\Delta^{-1}$, then $\bar{\Theta}$ is orthogonal and we can write $Q$ as
\begin{align}
\label{qcom}
Q&=\frac{\rho^2}{nL}\bar{\Theta}\Delta (\sum_{l=1}^L B_{l,0} \Delta^2 B_{l,0})\Delta \bar{\Theta}^\T \nonumber\\
&\succeq c \frac{\rho^2}{L} \bar{\Theta}\Delta (\sum_{l=1}^L B_{l,0}  B_{l,0})\Delta \bar{\Theta}^\T \nonumber\\
&\succeq c {\rho^2} \bar{\Theta}\Delta^2 \bar{\Theta}^\T \nonumber\\
&\succeq c {n\rho^2} \bar{\Theta} \bar{\Theta}^\T,
\end{align}
where we used Assumption \ref{sparse} in the first equality, Assumption \ref{balcom} in the first and last inequality, and Assumption \ref{rank} in the second inequality. (\ref{qcom}) implies that $Q$'s rank is $K$ and $\lambda_K(Q)\geq c n\rho^2$.

Next, we specify the upper bounds for $\mathcal I_1,\mathcal I_2$ and $\mathcal I_3$ and the corresponding condition in Theorem \ref{oneshot-theory} that $\mathcal I_1$ and $\mathcal I_2$ need to be smaller enough. Similar to the proof of Theorem \ref{centralbound}, we can show that with probability larger than $1-O(m(s+n)^{-\nu})$ for some positive constant $\nu$,
\begin{align}
\label{I1bound}
\small
\max_{i}\|\sum_{l\in S_i}(\overline{M}_l-Q_l)/|S_i|\|_2&\leq C\max \bigg\{\frac{(q\rho+1-q')^{1/2}n^{1/2}\rho \log^{1/2}(s+n)}{(q+q'-1) \sqrt{s}},\nonumber\\
&\quad\quad\frac{(q\rho+1-q')\log(s+n)}{(q+q'-1)^2\sqrt{s}},\, \frac{((q+q'-1)\rho+1-q')^2}{(q+q'-1)^2}\bigg\}.
\end{align}
When
\begin{equation}
\label{eq:rhofirstcondition}
\rho \lesssim \sqrt{\frac{\log(s+n)}{n}}\cdot \frac{(q\rho+1-q')^{1/2}}{q+q'-1},
\end{equation}
the second term in the RHS of (\ref{I1bound}) dominates the first one.  As a result,
\begin{align*}
\mathcal I_1 &:= \frac{\max_{i}\|\sum_{l\in S_i}\overline{M}_l/|S_i|-Q\|_2^2}{\lambda_{\min}^2(Q)}\\
&\lesssim \max \bigg\{\frac{(q\rho+1-q')^2}{(q+q'-1)^4}\cdot \frac{\log^2(s+n)}{n^2\rho^4s},\, \frac{((q+q'-1)\rho+1-q')^4}{(q+q'-1)^4n^2\rho^4}\bigg\}:=\mathcal E_1.
\end{align*}
It is then easy to see that
\begin{equation*}
n\sqrt{s}\cdot\frac{\rho^2(q+q'-1)^2}{q\rho+1-q'}\gtrsim {{\log (s+n)}}
\end{equation*}
and
\begin{equation*}
\frac{(q+q'-1)\rho+1-q'}{\rho (q+q'-1)}\lesssim n^{1/2},
\end{equation*}
are sufficient conditions for $\mathcal I_1 \lesssim 1$. For $\mathcal I_2:=\frac{\mathcal H^2(n,\rho, L)}{n^2\rho^4}$, it can be implied by Assumptions \ref{balcom}, \ref{sparse} and Definition \ref{balb} that, $\mathcal H \lesssim n\rho^2$, and thus $\mathcal I_2 =\mathcal E_2\lesssim 1$ naturally. For $\mathcal I_3$, we have by Theorem \ref{centralbound} and (\ref{eq:rhofirstcondition}) that,
\begin{align*}
\small
\mathcal I_3\lesssim\max \bigg\{\frac{(q\rho+1-q')\log(L+n)}{(q+q'-1)^2\sqrt{L}n\rho^2},\, \frac{((q+q'-1)\rho+1-q')^2}{(q+q'-1)^2n\rho^2}\bigg\}:=\mathcal E_3.
\end{align*}

Finally, noting (\ref{simplenote}), we obtain the results of Theorem \ref{sparsecor}.
 \QEDA

\subsection{Proof of Corollary \ref{corofcen}}
Following Theorem \ref{sparsecor}, we only need to compare $\mathcal E_1$ and $\mathcal E_2$ with $\mathcal E_3$. When
\begin{equation*}
m\lesssim  \frac{n\rho^2(q+q'-1)^2}{(q\rho+1-q')\log (s+n)}\cdot L^{1/2},
\end{equation*}
the first term in $\mathcal E_3$ dominates the first term in $\mathcal E_1$. The second term in $\mathcal E_3$ dominates the second term in $\mathcal E_1$ because the former is just the squared version of the latter and they are both of smaller order than 1.
And when
\begin{equation*}
\mathcal H\lesssim \max \bigg\{\frac{n^{1/2}\rho\,(q\rho+1-q')^{1/2}\log^{1/2}(L+n)}{(q+q'-1)L^{1/4}},\, \frac{n^{1/2}\rho\, ((q+q'-1)\rho +1-q')}{q+q'-1}\bigg\},
\end{equation*}
$\mathcal E_3$ dominates $\mathcal E_2$. Hence we obtain the results of Theorem \ref{sparsecor}.
\QEDA

\subsection{Proof of Theorem \ref{localmis}}
The general proving strategy is same with \citet{lei2015consistency,rohe2011spectral,zhang2022randomized}, among others. To fix ideas, we now recall and {introduce} some notation. $V$ denotes $K$ leading eigenvectors of $Q$ and $\tilde{V}$ denotes the $K$ leading eigenvectors produced by Algorithm \ref{Oneshotsvd}. Likewise, $\tilde{V}'={{\tilde{\Theta}}}{{X}}$ denotes the estimated eigenvectors by the $k$-means with $\tilde{V}$ being its input, where ${\tilde{\Theta}}$ is the estimated membership matrix and ${{X}}$ denotes the centriods.

First, we have that there exists $O\in \mathcal O_K$ such that
\begin{align*}
\phantomsection
\|{\tilde{V}'}- VO\|_{\tiny  F}^2&=\|{\tilde{V}'}-\tilde{V}+\tilde{V}-VO\|_{\tiny F}^2\nonumber \\
& \leq 2\|VO-\tilde{V}\|_{\tiny  F}^2+2\|\tilde{V}-VO\|_{\tiny  F}^2 \nonumber \\
&=4\|\tilde{V}-VO\|_{\tiny  F}^2\nonumber\\
&\leq C \|\tilde{V}-VO\|_2^2 \leq C \mathcal E_3^2,
\end{align*}
where $\mathcal E_3$ is defined in Theorem \ref{sparsecor}, the first inequality follows because we assume that ${\tilde{V}'}$ is the global solution minimum of the following $k$-means objective and $VO$ is a feasible solution by Lemma \ref{aSBM-eigen},
\begin{equation}
(\tilde{\Theta}, X)=\underset{{{\Theta}\in  \mathbb M_{n,K},{X}\in \mathbb R^{K\times K}}}{{\mathrm{arg\;min}}}\;\|{\Theta} {X}-\tilde{V}\|_{ \tiny F}^2,\nonumber
\end{equation}
and the last inequality follows from Theorem \ref{sparsecor} and the fact that $\min_{O\in\mathcal O_K}\|\tilde{V}-VO\|_2^2\lesssim \mathrm{ dist}(\tilde{V}, V)$.

Then, we begin to bound the fraction of misclassified nodes. Define
\begin{equation*}
\phantomsection
\mathcal M=\{i\in \{1,...,n\}:\; \|{\tilde{V}'}_{i\ast}- ({V}O)_{i\ast}\|_{\tiny \mathrm {2}}\geq \frac{\tau}{2}\},
\end{equation*}
with $$\tau=\min_{k\neq l}\sqrt{(n_{k})^{-1}+(n_{l})^{-1}}.$$
Then we can see obviously that
\begin{equation*}
\phantomsection
\frac{|\mathcal M|}{n}\leq  C \mathcal S^2 /(n\tau^2)\leq  C' \mathcal S^2.
\end{equation*}
where the last inequality follows from Assumption \ref{balcom}.

Finally, we show that the nodes outside $\mathcal M$ are correctly clustered to their underlying communities. Define $T_k:= G_k\backslash \mathcal M$, where $G_k$ denotes the set of nodes within the true community $k$. $T_k$ is not an empty set provided that $|\mathcal M|<n_k$ for any $k$ by our condition. Let $T=\cup _{k=1}^{K}T_k$. We will rightly see that the rows in $(VO)_{T\ast}$ has a one to one correspondence with those in ${\tilde{V}'}_{T\ast}$. On the one hand, for $i\in T_k$ and $j\in T_l$ with $l\neq k$,
${\tilde{V}'}_{i\ast}\neq {\tilde{V}'}_{j\ast}$, otherwise the following contradiction follows
\begin{align*}
\phantomsection
\tau\leq \|(VO)_{i\ast}-(VO)_{j\ast}\|_2
\leq \|(VO)_{i\ast}-{\tilde{V}'}_{i\ast}\|_2+\|(VO)_{j\ast}-{\tilde{V}'}_{j\ast}\|_2
<\frac{\tau}{2}+\frac{\tau}{2},
\end{align*}
where the last inequality is implied by the definition of $\mathcal M$ and the first inequality is implied by Lemma \ref{aSBM-eigen}
by noting $$Q=\frac{1}{L}\sum_lP_l^2=\frac{1}{L}\sum_l\Theta B_l \Theta^\T \Theta B_l \Theta^\T =\bar{\Theta} \Delta \frac{1}{L}\sum_l (B_l \Delta^2 B_l) \Delta\bar{\Theta}^\T:={\Theta} \overline{B} {\Theta}^\T,$$
where $\Delta={\diag}(\sqrt{n_1},...,\sqrt{n_{K}})$, $\bar{\Theta}:=\Theta \Delta^{-1}$, $\bar{\Theta}^\T \bar{\Theta}=\mathbb I$ and $\bar{B}$ is of full rank by our condition (see (\ref{qcom}) for explanations).
On the other hand, for $i,j\in T_k$,
${\tilde{V}'}_{i\ast}= {\tilde{V}'}_{j\ast}$, because otherwise ${\tilde{V}'}_{T\ast}$ has more than $K$ distinct rows which is contradict with the fact that the community size is $K$.

So far, we have arrived at the conclusion of Theorem \ref{localmis}.
\QEDA

\section{Auxiliary lemmas}
\phantomsection
\label{app::lemma}
We provide auxiliary lemmas that are used for proving the theorems in the paper.

\begin{definition}[Bernstein tail condition]
\label{bernstein}
A random variable $Y$ is said to be $(v,R)$-Bernstein if $\ME[|Y|^k]\leq \frac{v}{2}k!R^{k-2}$ for all integers $k\geq 2$.
\end{definition}

\begin{lemma}[Theorem 3 in \citet{lei2020bias}]
\label{lei-bound1}
 Let $X_1,...,X_L$ be a sequence of independent $n\times r$ random matrices with zero-mean independent entries being $(v_1, R_1)$-Bernstein. Let $H_1,...H_L$ be any sequence of $r\times n$ non-random matrices. Then for all $t>0$,
$$\MP(\|\sum_{l=1}^L X_lH_l\|_2\geq t)\leq 4n\exp\left(-\frac{t^2/2}{v_1n\|\sum_l H_l^\T H_l\|_2+R_1 \max_l\|H_l\|_{2,\infty} t}\right).$$
\end{lemma}\QEDA

\begin{lemma}[Theorem 4 in \citet{lei2020bias}]
\label{lei-bound2}
Let $X_1,X_2,...,X_L$ be independent $n\times n$ random symmetric matrix with independent diagonal and upper diagonal entries. For any $l\in[L]$ and any pair of $(i,j)(i\leq j)$, $X_{l,ij}$ is $(v_1, R_1)$-Bernstein, and $X_{l,ij}\tilde{X}_{l,ij}$ is $(v'_2, R'_2)$-Bernstein where $\tilde{X}_{l,ij}$ is an independent copy of $X_{l,ij}$. Then with probability larger than $1-O((n+L)^{-1})$,
\begin{align*}
\|\sum_l X_l^2-{\diag}(\sum_lX_l^2) \|_2\leq C&\Big[v_1 n \log (L+n)\sqrt{L} +\sqrt{v_1}R_1\sqrt{Ln}\log^{3/2}(L+n)\nonumber\\
&\quad\quad+\sqrt{v'_2} \sqrt{Ln}\log (L+n)+(R_1^2+R'_2)\log^2(L+n) \Big].
\end{align*}
Moreover, if the squared entry $X_{l,ij}^2$ is $(v_2, R_2)$-Bernstein, then with probability larger than $1-O((n+L)^{-1})$.
\begin{align*}
\|{\diag}(\sum_l X_l^2)-\mathbb E({\diag}(\sum_l X_l^2)) \|_2\leq C\left[\sqrt{v_2}\log(L+n)\sqrt{Ln}+R_2\log(L+n)\right].
\end{align*}
\end{lemma}\QEDA

\begin{lemma}[Davis-Kahan theorem]
	\label{lem:DK}
	Let $Q\in\mathbb R^{n\times n}$ be a rank $K$ symmetric matrix with smallest nonzero singular value $\gamma_n$. Let $M$ be any symmetric matrix and $\hat{U},U\in \mathbb R^{n\times K}$ be the $K$ leading eigenvectors of $M$ and $Q$, respectively. Then there exists a $K\times K$ orthogonal matrix $O$ such that
	\[
	 \|\hat{U}-UO\|_{F} \leq\frac{2\sqrt{2K}\|M-Q\|_2}{\gamma_n}.
	\]
\end{lemma}
\emph{Proof.} See \citet{lei2015consistency,chen2021spectral}, among others.
\QEDA

\begin{lemma}
\label{aSBM-eigen}
Define the population matrix $Q=\Theta \bar{B}\Theta^\intercal$. Suppose $\bar{B}$ is of full rank and denote the eigen-decomposition of $Q$ by ${V}_{n\times K}{\Sigma}_{K\times K}{V}^\intercal_{K\times n}$. Then, $V=\Theta X$ for some given matrix $X$. Specifically, for $\Theta_{i\ast}=\Theta_{j\ast}$, we have ${V}_{i\ast}={V}_{j\ast}$; while for $\Theta_{i\ast}\neq\Theta_{j\ast}$, we have $\|{V}_{i\ast}-{V}_{j\ast}\|_2=\sqrt{(n_{g_i})^{-1}+(n_{g_j})^{-1}}$.
\end{lemma}
\emph{Proof.} See \citet{lei2015consistency,guo2020randomized,zhang2022randomized}, among others.
\QEDA

\spacingset{0.9}

\bibliographystyle{chicago}
\bibliography{dpnetwork}

\end{document}